\documentclass[12pt,preprint]{aastex}
\usepackage{graphicx}
\usepackage{subfigure}
\usepackage{epsfig}

\newcommand{\f}[2]{\frac{#1}{#2}}
\newcommand{\g}{\gamma}
\def\be{\begin{equation}}
\def\ee{\end{equation}}
\def\ba{\begin{eqnarray}}
\def\ea{\end{eqnarray}}
\newcommand{\swift}{{\it Swift }}

\begin{document}

\title{Relation between the intrinsic and observed central engine activity time: implications for ultra-long GRBs}

\author{He Gao$^{1,2,3}$, Peter M\'esz\'aros$^{1,2,3}$}
\affil{$^1$Department of Astronomy and Astrophysics, Pennsylvania State University, 525 Davey Laboratory, University Park, PA 16802: hug18@psu.edu\\
$^2$Department of Physics, Pennsylvania State University, 525 Davey Laboratory, University Park, PA 16802\\
$^3$Center for Particle and Gravitational Astrophysics, Institute for Gravitation and the Cosmos, Pennsylvania State University, 525 Davey Laboratory, University Park, PA 16802
}

\begin{abstract}
The gamma-ray burst (GRB) central engine intrinsic activity time $T_{\rm ce}$ is usually described through either the $\gamma$-ray duration $T_{90}$ or through a generalized burst duration $t_{\rm burst}$ which includes both the $\gamma$-ray emission and (when present) an extended flaring X-ray plateau. Here, we define a more specific operational description of $T_{\rm ce}$, and within the framework of the internal-external shock model, we develop a numerical code to study the relationship between $T_{90}$ and $T_{\rm ce}$, as well as between $t_{\rm burst}$ and $T_{\rm ce}$, for different initial conditions. We find that when $T_{\rm ce}\lesssim 10^4$ s, late internal collisions or refreshed external collisions result in values of $T_{\rm 90}$ and $t_{\rm burst}$ larger than $T_{\rm ce}$, usually by factors of $2-3$. For $T_{\rm ce}\gtrsim 10^4$ s, the $t_{\rm burst}$ is always a good estimator for $T_{\rm ce}$, while $T_{90}$ can underpredict $T_{\rm ce}$ when the late central engine activity is moderate. We find a clear bimodal distribution for $T_{\rm ce}$, based on our simulations as well as on the observational data for $T_{90}$ and $t_{\rm burst}$. We suggest that $t_{\rm burst}$ is a reliable measure for defining ``ultra-long" GRBs. Bursts with $T_{90}$ of order $10^3$ s need not belong to a special population, while bursts with $t_{\rm burst} > 10^4$ s, where the late central engine activity is more moderate and shows up in X-rays, may represent a new population. These conclusions are insensitive to the initial conditions assumed in the models.
\end{abstract}  
  
\section{INTRODUCTION}

After decades of study, our knowledge of the progenitor for gamma-ray bursts (GRBs) is still limited due to the lack of direct observational constraints \citep[e.g.,][]{zhang11}. In the CGRO/BATSE era, people suggested that the observed temporal behavior of GRB prompt emission essentially reflects the temporal behavior of the central engine \citep{rees94,kobayashi97}, which might provide clues about the progenitor. GRBs were classified into two categories: long-duration, soft-spectrum class (LGRBs) and the short-duration, hard-spectrum class (SGRBs), based 
on the bimodal distribution of GRBs in the duration-hardness diagram
\citep{kouveliotou93}. Different type of progenitor were invoked for these two different class, i.e., core collapse from Wolf-Rayet star for LGRBs \citep{woosley93,paczynski98,macfadyen99,woosley06} and mergers of two compact stellar objects (NS-NS and NS-BH systems) for SGRBs \citep{paczynski86,eichler89,paczynski91,narayan92}. Later observational results, such as the host galaxies types for each class, the supernova association identification for nearby GRBs, the localization of the burst with respect to their host galaxy, etc \cite[][for a review]{berger14}, seems to support such interpretations for the progenitor.

Recently, a number of GRBs (GRBs 101225A, 111209A, 121027A and 130925A) have attracted attention due to their unusually long prompt duration  ($\sim$ hours instead of tens of seconds) compared to
typical GRBs \citep{levan14,gendre13,virgili13,stratta13}, where ``prompt" refers to the initial emission phase, before the afterglow proper. It is worth noting that these ``ultra-long" GRBs are not the only cases, similar ones were also seen historically in BATSE and Konus-Wind data \citep[e.g.][]{connaughton97,connaughton98,connaughton02,giblin02,nicastro04,levan05,pal'shin08}. In these references, the ``ultra-long" GRBs were defined based on their $\g$-ray duration $T_{90}$, when $T_{90}$ is comparable or even larger than $10^3$ s\footnote{Although no clear boundary line has been defined yet.}. Some authors \citep{gendre13,nakauchi13,levan14}  proposed that these ``ultra-long" GRBs may issue from a new type of progenitor stars with much larger radii, such as blue supergiants \citep{meszaros01,nakauchi13}, which could naturally explain their unusually long durations. 

However, it has been widely argued that the $\g$-ray duration is not a good reflection of the intrinsic central engine activity $T_{\rm ce}$, since \swift observations suggest that many GRBs have an extended central engine activity time, manifested through flares \citep{burrows05a,zhang06,margutti11} and extended shallow plateaus \citep{troja07,liang07} in the X-ray light curves following the MeV emission. Most recently, \cite{zhang14} performed a comprehensive study on a large sample of \swift GRBs, in which they redefined the burst duration as $t_{\rm burst}$, based on both $\gamma$-ray and the above X-ray light curve features. They found that within their ``good'' sample, 21.9\% GRBs have $t_{\rm burst}\gtrsim10^3$ s and 11.5\% GRBs have $t_{\rm burst}\gtrsim10^4$ s. They showed that GRBs with exceedingly long $t_{\rm burst}$ do not necessarily have unusual $T_{90}$ and those four traditional ``ultra-long" GRBs are not among the longest ones although they do have relatively long $t_{\rm burst}$ ($> 10^4$ s). They conclude that, in the sense of $t_{\rm burst}$, the existing evidence is inadequate to separate those traditional ``ultra-long" bursts from a new population. Interestingly, there is an apparent bimodal distribution of $t_{\rm burst}$ in their results, separating around $10^4$ s (see their Fig 5). Although they claimed some selection effects that may strongly affect the distribution, the effects only apply to the first component ($<10^4$ s). Therefore, based on $t_{\rm burst}$, the second component (distributed from $10^4\sim10^{6}$ s) might be a real population that points toward a new type of progenitors.

Compared to $T_{90}$, which refers purely to MeV-range emission, $t_{\rm burst}$ has the advantage of better reflecting $T_{\rm ce}$, especially when late activity becomes more moderate. However, late time features need not necessarily be related to late central engine activity, since they might be due to the late internal collisions or refreshed external collisions \citep{rees98,sari00} from early ejected shells. Therefore in some cases, $t_{\rm burst}$ might overestimate $T_{\rm ce}$. Before deciding whether $T_{90}$ or $t_{\rm burst}$ should be used to define ``ultra-long" bursts, or determining whether the ``ultra-long" GRBs indeed require a new type of progenitor, it is crucial to gain a better understanding of the real distribution of $T_{\rm ce}$.

Here we develop a numerical method to study the behavior of the ratios $T_{90}/T_{\rm ce}$ and $t_{\rm burst}/T_{\rm ce}$ within the framework of the internal plus external shock \citep{rees94,meszaros97,kobayashi97,maxham09,gao13}. We estimate the distribution of $T_{\rm ce}$ using Monte Carlo simulations based on our models and on the observational data on the $T_{90}$ distribution of all \swift GRBs\footnote{These data are collected from the website: http://swift.gsfc.nasa.gov/archive/} and the data on the $t_{\rm burst}$ distribution from \cite{zhang14}. To better understand the intrinsic features of the ratios $T_{90}/T_{\rm ce}$ and $t_{\rm burst}/T_{\rm ce}$, the values of $T_{90}$ and $t_{\rm burst}$ in the rest frame are essential. On the other hand, when estimating $T_{\rm ce}$ by comparing with observational results of $T_{90}$ and $T_{\rm burst}$, the redshift correction should be considered. The paper is structured as follows: in section 2, we describe our numerical method for estimating $T_{90}$ and $t_{\rm burst}$ in the framework of the internal plus external shock model. The simulation details for obtaining realizations of our method are illustrated in section 3. We present the results for $T_{90}/T_{\rm ce}$ and $t_{\rm burst}/T_{\rm ce}$ under different initial conditions, as well as the estimation for distribution of $T_{\rm ce}$ in section 4. We discuss our method and the results in Section 5.  For convenience, we collect the definition of the various timescales in Table 1. Note that henceforth, in the rest of the manuscript, the redshift correction term ($(1+z)^{-1}$) for the timescales in the observer frame is omitted in all relevant formulae, for the sake for brevity. 

\section{Method description}
\label{sec:modeldescription}

In the internal plus external shock model, a relativistic unsteady outflow is generated from the central engine within the ejection time $T_{\rm ej}$ \citep{rees94}, which produces highly time-variable shocks lying inside the external blast wave radius. For simplicity, the outflow can be represented as a succession of shells ejected with random values of the Lorentz factor, mass and width. Due to nonuniformities in the velocity, mechanical collisions between adjacent shells occur, and a certain fraction of the relative kinetic energy is converted into internal energy by the resulting internal shocks. The internal energy thus generated is then radiated via synchrotron or inverse Compoton scattering giving rise to the prompt $\g$-ray emission. We will call  such collisions ``effective collisions" when they produce detectable radiation. There are two timescales in the lab frame to describe the collision history, $T_{\rm col,lab}$ for the last mechanical collision, and $T_{\rm eff,lab}$ for the last effective collision, where $T_{\rm col,lab}\geq T_{\rm eff,lab}$. Their values in the observer frame are denoted $T_{\rm col}$ and $T_{\rm eff}$. Besides internal collisions, the outermost shell will also interact with the external ambient medium, sweeping up an increasing amount of matter, which leads to its eventually slowing down. The trailing shells may collide with each other before interacting with the leading shell, but it is also possible that they collide againts the leading decelerating shell (or shells) before they have undergone internal collisions, which confines the possible values of $T_{\rm col}$ and $T_{\rm eff}$. 

In previous numerical studies \citep{kobayashi97,maxham09} $T_{\rm col}$ has been usually taken to be the prompt emission duration $T_{90}$ (see section 5 for a detailed discussion about the difference between this work and previous works). As a better definition, we propose to use $T_{\rm eff}$ to represent $T_{90}$, taking also into account the detector properties for the detectability. Specifically, we use three criteria to define ``effective internal collisions for $\g$-rays" : firstly the Lorentz factor ratio between two shells should be nominally $\gtrsim 2$ \citep{rees94}\footnote{This is a somewhat ad-hoc criterion, which in a more careful analysis could be loosened, but which qualitatively serves our imediate purposes here.}; secondly the peak radiation luminosity should be larger than the $\g$-ray detector sensitivity $L_{\rm det,\g}$; and finally, the peak photon energy $E_p$ should be within the $\g$-ray detector energy band. The final estimated prompt emission duration is denoted as $T_{\rm pr,obs}$, where $T_{\rm pr,obs}\leq T_{\rm eff}$.

When the leading shell sweeping up enough matter, its kinetic energy is converted into afterglow emission via external shock \citep{meszaros97}. Although the afterglow timescale is no longer connected with $T_{\rm ej}$, if ``effective internal collisions for X-rays"  happen after the onset of the afterglow emission, or if energetic late shells collide onto the external blastwave (effective refreshed collisions), then we expect observable signatures such as flares or plateaus, which may be detected superposed on the external shock afterglow signals (at time $T_{\rm af}$). In this case, the central engine timescale inferred from the observations should be $T_{\rm {ce,obs}}={\rm max}(T_{\rm pr,obs},~T_{\rm af})$. In \cite{zhang14} the X-ray light curve was fitted with a multi-segment broken power-law and $T_{\rm burst}$ was identified as the maximum of the steep (decay slope steeper than $-3$) to shallow transition time. In our model, $T_{\rm af}$ is defined as the time when the last effective internal collision for X-rays or the last effective refreshed shock collision happens. The criterion for an effective internal collision for X-rays is similar to that for $\g$-rays, but the X-ray detector properties are used. For an effective refreshed collision, the collision needs to be violent, or else the injection energy needs to be comparable to impulsive energy in the initial blastwave if the collision is mild\footnote{For a violent collision, a significant X-ray/optical flare would be produced, which could be easy to identify. For a mild collision, a shallow decay phase may appear in the afterglow light curve; however, to make it detectable, one would need $\delta t_{\rm obs}/t_{\rm col,obs}=E_2/E_1$ to be of order unity (see section \ref{sec:refreshed shock}), i.e., the injection energy should be  comparable to the initial blastwave energy.  The criteria for violent collisions are defined later in section 3.1.2. } \citep{zhang02}. Although the definition of  $T_{\rm ce,obs}$ is not fully equivalent with $t_{\rm burst}$, for a large enough sample, it should be adequate to represent the overall distribution feature of $t_{\rm burst}$.

In principle, given the initial conditions, i.e., the Lorentz factor, mass and width of the shells, the ejection time intervals between shells and the ambient medium density, the dynamical processes in the system can be precisely described through the numerical simulations. This allows us to study the expected values of the ratios $T_{\rm pr,obs}/T_{\rm ej}$ and $T_{\rm ce,obs}/T_{\rm ej}$ in our model, which essentially represent the ratios $T_{90}/T_{\rm ce}$ and $t_{\rm burst}/T_{\rm ce}$ between the observed and the model-intrinsic features. These, in turn, provide an estimate of the intrinsic distributions of the $T_{\rm ce}$, based on the observed values of $T_{90}$ and $t_{\rm burst}$. 

\section{Simulation Details}
\label{sec:sim}

\subsection{Two Shell Interaction}

The treatment of the two shell interaction is the basic element of this simulation. Most of the interactions are internal collisions between two adjacent shells. Only the interaction between the first and second shell should be treated differently when the first shell enters the deceleration phase, which is referred to as a refreshed collision \citep{rees98,sari00}. 

\subsubsection{Internal Collision}
\label{sec:internal shock}

Consider a rapid shell (at $R_r$) with initial width $l_{r}$, Lorentz factor $\g_r$ and mass $m_r$, chasing a slower shell (at $R_s$), which has an initial width $l_{s}$, Lorentz factor $\g_s$ and mass $m_s$. In the lab frame, the shells will collide after
\begin{eqnarray}
t_{\rm{col}}=\frac{R_s-R_r}{(\beta_r-\beta_s)c},
\end{eqnarray}
at 
\begin{eqnarray}
R_{\rm{col}}=\frac{\beta_r R_s-\beta_s R_r}{(\beta_r-\beta_s)}.
\end{eqnarray}
For simplicity, we assume the system behaves like an inelastic collision, so the two shells will merge to form a single new shell of width $l_{m}$, Lorentz factor $\g_m$ and mass $m_m$. Using conservation of mass, energy and momentum, we can estimate the mass and Lorentz factor of the merged shell as
\begin{eqnarray}
m_{m}=m_r+m_s
\end{eqnarray}
\begin{eqnarray}
\gamma _{m}\simeq \sqrt{\frac{m_{r}\gamma _{r}+m_{s}\gamma _{s}}{%
m_{r}/\gamma _{r}+m_{s}/\gamma _{s}}},  \label{gammam}
\end{eqnarray}
Based on a detailed hydrodynamic calculation, \cite{kobayashi97} proposed to estimate the width of merged shell as
\begin{equation}
l_{m}=l_{s}\frac{\beta _{fs}-\beta _{m}}{\beta _{fs}-\beta _{s}}+l_{r}\frac{%
\beta _{m}-\beta _{rs}}{\beta _{r}-\beta _{rs}}.
\end{equation}
where $\beta _{fs}$ is the speed of the forward shock propagating into the slow shell and $\beta_{rs}$ is the speed of the reverse shock propagating into the fast shell. Their corresponding Lorentz factors are
\begin{equation}
\gamma _{fs}\simeq \gamma _{m}\sqrt{\left( 1+\frac{2\gamma _{m}}{\gamma _{s}}%
\right) /\left( 2+\frac{\gamma _{m}}{\gamma _{s}}\right) },\ \ \ \gamma
_{rs}\simeq \gamma _{m}\sqrt{\left( 1+\frac{2\gamma _{m}}{\gamma _{r}}%
\right) /\left( 2+\frac{\gamma _{m}}{\gamma _{r}}\right) }.
\end{equation}
The internal energy of the merged shell is the difference between the kinetic energy before and after the collision: 
\begin{equation}
E_{\rm{int}}=m_{r}c^{2}(\gamma _{r}-\gamma _{m})+m_{s}c^{2}(\gamma _{s}-\gamma
_{m}).
\end{equation}
We define a free parameter $f$ as the fraction of the internal energy emitted in the detector energy band. 
The emission time scale (in the lab frame) can be estimated as the time taken by the reverse shock to cross the rapid shell: 
\begin{equation}
\label{deltat}
\delta t_{e}=\Delta_{r}/c(\beta _{r}-\beta _{rs}).
\end{equation}
where $\Delta_{r}$ is the width of the rapid shell, and 
\begin{equation}
\Delta_r  = \left\{
\begin{array}{l l}
  \l_r, & \quad R <R_{s,r} \\
  \f{R}{\gamma_r^2}, &\quad R > R_{s,r} .\\ \end{array} \right.\
\label{shellspreading}
\end{equation}
where  $R_{s,r}\sim \gamma_r^2 l_r$ is the spreading radius of the rapid shell.
The peak luminosity for this collision can thus be expressed as 
\begin{eqnarray}
L_p=E_{\rm{int}}fc(\beta _{r}-\beta _{rs})/\Delta_{r}
\end{eqnarray}
We assume that the energy $E_p$ of the spectral peak of the radiation can be estimated through the Amati relation, which is an empirical relation between the isotropic emission energy $E_{\rm iso}$ and $E_p$, generally valid among GRBs \citep{amati02,krimm09} and also within a single burst \citep{liang04,ghirlanda09}.
Here, we will assume the validity of this correlation for all effective internal collisions, and apply the relation $E_{\rm iso}=f E_{\rm int}$ to estimate $E_p$ as \citep{maxham09}
\begin{equation}
E_{p} =\frac{100 ~{\rm keV}}{1+z} \left( \f{E_{\rm iso}}{10^{52}~{\rm erg}}
\right) ^{1/2}.
\label{Amati}
\end{equation}
In this work, the detector sensitivities adopted are those of the \swift BAT and XRT instruments. For BAT, the detector threshold is taken to be $L_{\rm det,\g}=10^{-8}~ \rm{erg~s^{-1}~cm^{-2}}$, and the energy band is $15 - 350 \rm ~{keV}$ \citep{barthelmy05,maxham09}. For XRT, the sensitivity curve adopted is a broken power-law shape, with $\propto t^{-1}$ early on, which breaks to $\propto t^{-1/2}$ when $F_{\nu} = 2.0 \times 10^{-15}~ \rm erg~ cm^{-2} ~s^{-1}$ at $t=105~\rm s$. The energy band is $0.2-10 ~\rm{keV}$ \citep{burrows05b,moretti09}. These, together with the criteria discussed in section \ref{sec:modeldescription}, are used to label a collision as an effective collision for $\g$-rays, an effective collision for X-rays, or as not effective.

After the collision, the remaining internal energy $(1-f)*E_{\rm int}$ converts back into kinetic energy. In the following simulations, instead of considering the reacceleration details, we simply correct the bulk Lorentz factor of the merged shell to account for this energy conversion. We also tested situations where this effect is neglected, and it turned out that the final results were not affected. This could be due to the fact that the energy conversion efficiency for an individual internal collision is typically low, only (1-10)\%.

\subsubsection{Refreshed Shock Collisions}
\label{sec:refreshed shock}

The outermost shell will eventually slow down due to an external shock as it sweeps up the ambient medium. During the initial interaction, a pair of shocks (forward and reverse) propagate into the ambient medium and into the shell, respectively \cite[][]{saripiran95}. After the reverse shock crosses the shell, the blastwave enters a self-similar phase described by the Blandford– McKee self-similar solution \citep{blandford76}. The  deceleration radius $R_{\rm{dec}}$ beyond which the inertia from the circumburst medium is large enough and the Lorentz factor of the blastwave starts to decrease as a power law with radius is, for a constant density external medium $n_0$, given by
\begin{eqnarray}
R_{\rm{dec}}=\left(\frac{3E}{2\pi n_0 m_p c^2 \g_0^2}\right)^{1/3}
\label{eq:rdec}
\end{eqnarray}
where $E$ and $\g_0$ are the initial kinetic energy and Lorentz factor of the outermost shell. Entering  the self-similar phase, the dynamics of the blast wave in the constant energy regime can be described as \citep{blandford76}
\begin{eqnarray}
\gamma=\left(\frac{17E}{4^{8}\pi
n_0m_pc^{5}t^{3}}\right)^{1/8},~~~~~~~~R=\left(\frac{17Et}{\pi
n_0m_pc}\right)^{1/4},
\label{BM}
\end{eqnarray}
where $t$ is measured in the observer frame.

When the outermost shell is inside $R_{\rm{dec}}$, its collision with the second outermost shell can be taken as an internal collision (see section \ref{sec:internal shock}). Otherwise, the collision should be treated as a refreshed collision with the external shock. We denote the first outer shell with subscript $1$ and the second outer shell with $2$. The refreshed collision will happen at 
\begin{equation}
R_{\rm{col}}  \approx \left\{
\begin{array}{l l}
  R_{\rm{1}}+\left[8\gamma_{\rm{1}}^2(R_{\rm{1}}-R_{\rm{2}})R_{\rm{1}}^3\right]^{1/4}, & \quad \g_1<\g_2\\
  R_{\rm{1}}\left(\frac{\gamma_{\rm{2}}}{\gamma_{\rm{1}}}\right)^{-2/3}+\left[8\gamma_{\rm{1}}^2(R_{\rm{1}}-R_{\rm{2}})R_{\rm{1}}^3\right]^{1/4}, &\quad \g_1>\g_2 \\ \end{array} \right.\
\label{shellspreading}
\end{equation}
and the collision time is $t_{\rm{col}}=(R_{\rm{col}}-R_2)/\beta_2 c$. The detailed dynamics of such collisionis is complicated, involving three shocks and several distinct dynamical stages \citep{kumar00,zhang02}. For the purposes of this study, we adopt the following simple treatment: 1) after the complicated collision process and the relaxation stage (at $R_f$), the merged shell evolves with an initial total energy $E_1+E_2$ and an initial Lorentz factor $\rm{max} (\g_1, \g_2)$; 2)  between $R_{\rm{col}}$ and $R_f$, the shell 2 was assumed to inject its energy into shell 1 with a constant luminosity, which gives $R_f=R_{\rm{col}}(1+E_2/E_1)^{1/2}$. Therefore, the duration of the emission in the observer frame can be estimated as 
 \begin{eqnarray}
 \label{deltat:ref}
 \delta t_{\rm obs}=t_{\rm col,obs}E_2/E_1
 \end{eqnarray}
where $t_{\rm col,obs}$ is the relevant value for $R_{\rm{col}}$ in the observer frame. 

The collision can be classified as either violent or mild, depending on whether a strong shock 
forms at the discontinuity between two colliding shells, which requires two criteria: 
1) the injected shell should move supersonically with respect to the leading shell; 2) the injected 
shell is energetic enough to further heat up the leading shell. Quantitatively the criteria read 
(noticing that the sound speed $c_{s} \simeq c/\sqrt{3}$ for the leading shell):
\begin{eqnarray}
&&\g_{12}\geq 1.22 \nonumber\\
&&(4\g_{12}+3)(\g_{12}-1)>\frac{4E_1}{E_2}\left[\rm{min} \left(1,\frac{R}{R_{\rm{s,2}}}\right)\right]^{-1},
\label{eq:violent}
\end{eqnarray}
where $\g_{12}=1/2(\g_1/\g_2+\g_2/\g_1)$ is the relative Lorentz factor between shells 1 and 2, and $R_{\rm{s,2}}$ is the spreading radius of shell 2 \citep{zhang02}. We label the violent collisions or mild collisions with $E_2/E_1 \geq 1$ as effective refreshed collisions.

\subsection{The Multiple Shell Model}

Consider an outflow consisting of $N$ shells. We assign an index $i$, ($i=1,N$)
to each shell according to the order of the emission from the inner engine.
Each shell is characterized by four variables: a Lorentz factor $\gamma _{i}$,
a mass $m_{i}$, a width $l_{i}$ and the radius $R_{i}$ \footnote {For specific simulations, 
$R_{i}$ is calculated as $R_{i}=r_0+\beta_i c (\sum_i l_i/c + \sum_i\Delta t_{ej,i})$, where 
$\Delta t_{ej}$ is the ejection time interval between shells, $r_0$ is the central engine radius, 
whose value does not affect the final results as long as it is smaller than $c \Delta t_{ej,i}$. 
In this work, we adopt $r_0=10^7 \rm {cm}$.} . At this initial stage, the initial lab frame time 
is set to $t_{\rm lab,0}$. 

There will be $N-1$ collisions between different shells and eventually only one shell will be left, considering the confinement imposed by the external shock blastwave. For the $n$th collision (at $t_{\rm lab,n}$), we first find out all the groups of adjacent shells ordered with decreasing values of the Lorentz factor. We denote with an index $j$ for the $j$th group and with an index $s$ for the shells in each group ($s=1$ has the largest Lorentz factor ). With our results from \S \ref{sec:internal shock}, we can estimate the collision time $t_{\rm{j,s}}$ for the $s$ and $s+1$ shells in the $j$th  group, and find the pair with the shortest $t_{\rm{j,s}}$ among all the groups, assigning it the label $t_{\rm{next}}^{n}$. With the information of ($j$,$s$), we can find the shell index of the next expected collision pair ($h$,$h+1$). Note that one needs to check whether the outermost shell is already in the deceleration phase or not. If so, the outermost shell and the second outer shell should always be treated as the $j+1$ group. 

We then rearrange the shells as follows: for $i<=h$, each shell moves
from its earlier position $R_{i}(t_{\rm lab,n-1})$ to 
\begin{equation}
R_{i}(t_{\rm lab,n})=R_{i}(t_{\rm lab,n-1})+c\beta _{i}t_{\rm{next}}^{n}.
\end{equation}
The $i<h$ shells keep their other properties $\gamma _{i}$, $m_{i}$ and $l_{i}$, while $i=h$ is the new merged shell with the new $\gamma _{h}$, $m_{h}$ and $l_{h}$ calculated based on section \ref{sec:internal shock} (or section \ref{sec:refreshed shock} if the refreshed shock happen first). We delete the $h+1$ th shell. For $i>h+1$, each shell's index is reduced by one, and has a new position
\begin{equation}
R_{i-1}(t_{\rm lab,n})\equiv R_{i}(t_{\rm lab,n-1})+c\beta _{i}t_{\rm{next}}^{n}.
\end{equation}
They also keep their other properties. We then return to the calculation of the $(n+1)$ th step until all the shells have merged to form a single external blastwave, where the simulation stops. 

In the lab frame, the $n$th collision would happen at
\begin{eqnarray}
t_{\rm lab,n}=t_{\rm lab,0}+\sum_{1}^{n} t_{\rm{next}}^{n}.
\end{eqnarray}
For an observer at a luminosity distance $D_L$ from the central engine, the radiation from this collision would start to be detected at
\begin{equation}
t_{n}=(D_L-R_{\rm col,n})/c+t_{\rm lab,n}.
\end{equation}
At the end of the simulation,  we set the minimum of $t_n$ to be the origin of the observer time, since this is when the source triggered the detector. We thus have
\begin{equation}
t_{\rm obs,n}=t_n-\rm{min}(t_1,t_2,...t_{N-1}).
\end{equation}
For internal collisions, the observer-frame emission duration $\delta t_{\rm obs,n}$ is estimated as $\delta t_{\rm e,n}/2\g_m^2$, where $\g_m$ is the merged Lorentz factor for that collision. For refreshed collisions, we take $\delta t_{\rm obs,n}=t_{\rm obs,n}\times (E_2/E_1)$, where $E_1$ is the blastwave energy and $E_2$ is the subsequent injected energy for that collision. We search for the last effective internal collision for $\g$-rays (labelled as $n\gamma$), the last effective internal collision for X-rays (labelled as $nX$), and the last effective refreshed collision (labelled as $nR$) in the observer frame, leading to
\begin{eqnarray}
T_{\rm pr,obs}&=&t_{\rm obs,n\gamma}+\delta t_{\rm obs,n\gamma}\backsimeq T_{90},\nonumber\\
T_{\rm ce,obs}&=&{\rm max}~(t_{\rm obs,n\gamma}+\delta t_{\rm obs,n\gamma},~t_{\rm obs,nX}+\delta t_{\rm obs,nX},~t_{\rm obs,nR}+\delta t_{\rm obs,nR})\backsimeq t_{\rm burst}
\end{eqnarray}

\subsection{Initial conditions} 
\label{sec:initcond}

For each simulation, the initial conditions include the Lorentz factor, mass and width of the shells ($\g$, $m$ and $l$), the ejection time intervals between shells ($\Delta t_{\rm ej}$), the ambient medium density and the redshift of the source. In this work, we assume a constant density for the ambient medium, with $n=1~\rm cm^{-3}$. For each simulation, the source redshift is simulated based on the observed $z$ distribution of the observed GRBs\footnote{The data was collected from an online catalog listed at http://lyra.berkeley.edu/grbox/grbox.php.} (see Figure 1(b) in \cite{gao14}). Other parameters are determined by the central engine, and these are very uncertain. We tested different distribution functions for each parameter, especially for the Lorentz factor of the shells and the shell ejection time interval, which are more crucial for determining the observable timescales. We allow for one quiescent period in the central engine activity, that is, the shells may be ejected within one episode, or within two episodes separated by $T_{\rm quie}$. As a nominal case, we assume that 100 shells are ejected in each episode. The second episode can either share the same set of initial conditions as the first one, or it can be less energetic by assigning a lower mass to each shell. Details of the combinations of initial conditions are collected in Table 2, numbered as different models. Here ${\rm RAN} ~(\tilde{a},\tilde{b})$ denotes a uniform distribution from $\tilde{a}$ to $\tilde{b}$; ${\rm PL}~(\tilde{a},\tilde{b},\tilde{c})$ denotes a power-law distribution from $\tilde{a}$ to $\tilde{b}$ with index $\tilde{c}$; ${\rm GAUSS}~(\tilde{a},\tilde{b})$ denotes a Gaussian distribution with a mean $\tilde{a}$ and a standard deviation $\tilde{b}$. 

The reason for selecting these models can be briefly summarized as follows: for one episode 
ejection, we assume that the plausible shell ejection time interval are distributed as ${\rm RAN}(0.5,2.5)$, 
giving a total ejection timescale of $\sim 100 \rm s$ for 100 shells. Concerning the Lorentz factor 
of the shells, we assume the most plausible parameters for each of the three distribution functions, 
e.g. ${\rm RAN}(50,500)$, ${\rm PL}(100,500,-0.5)$ and ${\rm GAUSS}(300,100)$, to ensure a mean value of the Lorentz 
factor of $\sim 300$. Furthermore, we assume the mass of the shells is distributed with a uniform 
distribution in log space $10^{27} \sim 10^{29}$ g, giving a total energy of $10^{52}\sim10^{54} 
\rm{erg}$. These combined assumptions are used to propose what we consider to be the three most 
plausible models for the initial conditions of the GRB central engine, e.g. model 1 (RAN-type), 
7 (PL-type) and 9 (GAUSS-type). To account for the variability of the central engine properties, 
other specific models are also considered. For instance, model 2 is for cases when the shells have 
higher average Lorentz factors; models 3, 8 and 10 are for cases when the total ejection timescale 
is $\sim 1000$ s; models 11-22 are for two episode ejection cases, where the second episode has 
the same or is less energetic compared to the first one, and the separation timescale between the
two ejections is $10^3$ s or $10^4$ s.

Note that for given model, the central engine activity time is calculated as 
\begin{eqnarray}
T_{\rm ej}\equiv T_{\rm ce} =\sum_i l_i/c + \sum_i\Delta t_{ej,i} +T_{\rm quie}
\end{eqnarray}

\section{Results}
\label{sec:results}

\subsection{Summary of the Numerical Results}
\label{sec:sumres}

For each model, we run the simulation 500 times \footnote{The number of runs for each
simulation is determined by balancing the computation time consumption and the resulting convergence.}, 
and we analyze the thus obtained distributions of $T_{\rm pr,obs}/T_{\rm ej}$ and 
$T_{\rm ce,obs}/T_{\rm ej}$ in the rest frame. In Figure 1 we plot the results for selected models 
which are relevant for reflecting the main conclusions, which can be summarized as follows 
(for easy identification, Table 2 summarizes the corresponding subfigure numbers for the different 
models): 

\begin{itemize}

\item When the initial $\g$'s of the shells are distributed from 500 to 1000, compared to the case when they are distributed from 50 to 500, both $T_{\rm pr,obs}/T_{\rm ej}$ and $T_{\rm ce,obs}/T_{\rm ej}$ have a much narrower distribution around unity in the former case, due to a stronger confinement from the external blast wave. 

\item Different distribution functions for the initial $\g$'s of shells affect only moderately the results. The RAN and GAUSS functions give very similar results as long as they have similar distribution ranges. The results for ${\rm PL} (\tilde{a},\tilde{b},\tilde{c})$ are sensitive to the lower ending $\tilde{a}$ and the index $\tilde{c}$ of the distribution. Smaller $\tilde{a}$ or $\tilde{c}$ tends to give more spread-out distributions and larger maximum values for both $T_{\rm pr,obs}/T_{\rm ej}$ and $T_{\rm ce,obs}/T_{\rm ej}$.

\item When $T_{\rm ej}$ is of order of 100 s,  the $T_{\rm pr,obs}/T_{\rm ej}$ distribution has a sharp peak at unity, and a Gaussian-shape spreading from 0 to 2, depending on specific models. $T_{\rm ce,obs}/T_{\rm ej}$ behaves as a FRED-like (fast rise exponential decay) distribution from 1 to less than 10, also depending on specific models. 

\item For one-episode injection cases, when $T_{\rm ej}$ is of order of 100 s, $T_{\rm pr,obs}/T_{\rm ej}$ has an upper limit of order of 10, and $T_{\rm ce,obs}/T_{\rm ej}$ has an upper limit of order of 100 (for most cases, the upper limit is also of order 10, except when $\g$ has a PL distribution starting from 50). On the other hand, when $T_{\rm ej}$ is of order 1000 s, $T_{\rm pr,obs}/T_{\rm ej}$ has a very narrow distribution within $0-2$ and $T_{\rm ce,obs}/T_{\rm ej}$ also has a relatively narrow distribution with an upper limit smaller than 10. These results indicate that for one-episode injection, $T_{\rm pr,obs}$ has an upper limit of order 1000 s, and $T_{\rm ce,obs}$ has an upper limit of order $10^4$ s. These upper limits are set by the confinement of the blast wave and are model independent. 

\item For two-episode injection cases, when the second episode is similar to the first one, $T_{\rm pr,obs}/T_{\rm ej}$ has a very narrow distribution around unity, with some rare exceptions distributed around 0.1 (due to the high redshift). Also $T_{\rm ce,obs}/T_{\rm ej}$ is mainly concentrated around unity, spreading up to $2-3$. When the second episode is less energetic, 
$T_{\rm pr,obs}$ is essentially determined by the first episode, which has an upper limit of order 1000 s, and the $T_{\rm ce,obs}/T_{\rm ej}$ is distributed narrowly around unity. In short, $T_{\rm ce,obs}$ reflects well the time $T_{\rm ej}$ during which the late central engine is active, whereas $T_{\rm pr,obs}$  can be much smaller than $T_{\rm ej}$, if the  later injection is less energetic. 

\item The overall internal collision efficiency (total internal energy created from internal collision divided by total energy of ejected shells) is distributed with a Gaussian shape. The peak value for the RAN and GAUSS models are around $30\% - 40\%$, while the PL models are less efficient, with a peak around $15\%-30\%$.
\end{itemize}

\subsection{Distribution of the Intrinsic Central Engine Activity Timescale}
\label{sec:distres}

In the above simulations, $T_{\rm ce}$ ($T_{\rm ej}$) are concentrated in a very narrow range, for a given model. In reality, $T_{\rm ce}$ should have a much wider distribution, in order to produce the observed distribution profiles of $T_{90}$ ($T_{\rm pr,obs}$) and $t_{\rm burst}$ ($T_{\rm ce,obs}$). The observed distributions of $T_{90}$ of all \swift bursts, and of $t_{\rm burst}$ in the good sample of \cite{zhang14}, are plotted in Figure 2. With a better understanding about the relation between $T_{90}$ and $T_{\rm ce}$, and also between $t_{\rm burst}$ and $T_{\rm ce}$, we can now provide a better estimate for $T_{\rm ce}$. 

Since $t_{\rm burst}$ becomes very close to $T_{\rm ce}$ when $T_{\rm ce}\gtrsim10^{4}$ s, the second component in the $t_{\rm burst}$ distribution (from $10^{4}\sim10^{6}$ s) should show a similar structure in the $T_{\rm ce}$ distribution. If we treat this late time activity as a second episode injection, it must be less energetic compared to the first one, since the $T_{90}$ distribution lacks a corresponding excess. Based on the same argument, one continuous long episode is also ruled out. On the other hand, for the majority of cases when $T_{\rm ce}\lesssim10^{3}$ s, both $t_{\rm burst}$ and $T_{\rm 90}$ would be comparable to, or a slight overestimate of $T_{\rm ce}$. Since both $T_{90}$ and the first component for $t_{\rm burst}$ has a cutoff around $10^{3}$ s, another component for $T_{\rm ce}$ is required, whose distribution profile should be similar to $T_{\rm 90}$ and the first component of $t_{\rm burst}$, but with an overall shift.

As a good representation of the observed distributions (the dashed lined histograms in Figure 2, we propose a Gaussian distribution for the first component of $T_{\rm ce}$ and a Gaussian distribution in log space for the second one (see Figure 2a). The population ratio between the two components is $5:1$. We apply the simulation results of three plausible models, e.g., model 1 (RAN-type), 7 (PL-type) and 9 (GAUSS-type) to the first component. Moreover, since the second component should be less energetic, we correspondingly apply the results of model 13 (RAN-type), 17 (PL-type) and 22 (GAUSS-type) to the second component.  In principle, a distribution test could be used here to find out the best parameters for the $T_{\rm ce}$ distribution. Due to an incomplete understanding of the real $T_{\rm ce}$ distribution profile, and an incomplete unnderstanding of relevant  selection effects \citep{zhang14}, we only show in Figure 2 an example of a $T_{\rm ce}$ distribution that gives reasonable results for both $T_{90}$ and $t_{\rm burst}$, instead of using a distribution test to find out the best parameters. 

In this example, the first component is a ${\rm GAUSS} (60,80)$ and the second one is a ${\rm GAUSS}(4.5,0.5)$ in log space.
The results are shown in Figure 2. We find that the proposed $T_{\rm ce}$ distribution provides a good match to the observational results, and it is largely insensitive to the initial conditions of the internal-external shock models used. If our estimates are indeed as robust as they appear, the following implications can be inferred:

1) For the first component, our model $T_{\rm ce}$ is smaller than 300 s within $3\sigma$, and appears adequate to account for the rare observed tails of $T_{90}$ which extend to over $10^3$ s. This suggests that bursts with $T_{90}$ of order of $10^3$ s need not belong to a special population, and their relevant intrinsic central engine activity time may be even smaller than 300 s.

2) Unlike the distribution structure of $t_{\rm burst}$, there is an obvious gap between the two components of $T_{\rm ce}$, indicating a more clearly bimodal distribution. In this sense, bursts with $t_{\rm burst} > 10^4$ s might indeed belong to a new population, and it is more reliable to use $t_{\rm burst}$ to define ``ultra-long" GRBs;

3) For the``ultra-long" GRBs defined by $t_{\rm burst}$, the late central engine activity tends to be much more moderate than the early activity.

\section{Discussion and Conclusions}

Within the framework of the internal plus external shock model, we have developed a numerical code to study in greater detail the relationship between $T_{90}$ and $T_{\rm ce}$, as well as between $t_{\rm burst}$ and $T_{\rm ce}$, for a range of different initial conditions. 

In the literature, previous numerical simulations for internal shocks have been done, but aimed at different problems, e.g.,  testing the ability of internal shock models to produce the temporal variability \citep{kobayashi97}, studying the power density spectrum of GRB lightcurves in internal shock model \citep{panaitescu99}, studying the detailed radiation properties of internal shock models \citep{daigne98} and modeling the X-ray flares \citep{maxham09}. The main difference between our present study and the previous ones is that we use  both the internal collision shock properties and the real detector sensitivities to estimate the prompt duration; we consider the confinement imposed by the external blastwave to the late residual collisions\footnote{Note that the detector sensitivity and the blast wave refreshing by subsequent collisions were also considered in \cite{maxham09}, when they estimate the observable features of X-ray flares.}; for the first time, we use a  detailed criterion to define effective refreshed collisions to calculate the observable central engine activity time; for the first time, we randomly generate all the initial condition parameters with many different distribution functions, intercomparing the results. Finally, 
since the real detector sensitivities are used, it is essential to assign a reasonable redshift to 
each simulated source for internal consistency. In this work, we rely on the actual observational GRB redshift distribution 
information rather than assuming one average redshift for all the sources, as has been done in some 
previous works. Such a treatment is also important when connecting intrinsic timescales in the rest 
frame with the corresponding values in the observer frame.

Some caveats about our numerical method and how these might affect the results are:

1) At the shell injection stage, we assume that the shell is released with relativistic speed, rather than considering the acceleration process for the shells. This should not affect our results since the acceleration phase should finish well below the internal shock radius.

2) We did not consider the details of the internal shock radiation. Here to estimate $L_p$ we assumed that $f=0.5$, i.e., $50\%$ of the energy dissipated during collisions goes into radiation, and we used the Amati relation to calculate the spectral peak energy $E_p$. This value is adopted for keeping the overall radiation efficiency $\eta_{\g}=E_{\g}/(E_{\g}+E_{K})$ at $10\%-20\%$, where $E_{\g}$ is the isotropic gamma-ray energy and $E_{K}$ is the isotropic kinetic energy \citep{zhang07}. To justify this assumption, we re-simulated  all the results by using $f=1$ (see pink lines in Figures 1). It turns out the results are almost independent on the $f$ value, especially for $T_{\rm ce,obs}$. The issue of whether $f=0.5$ (or $f=1$) might be too large in the sense of the electron equipartition energy fraction $\epsilon_e$ would lead to the so-called ``efficiency problem" for the internal shock model \cite[][and reference therein]{granot06}, which is out of the scope of this paper. However, it is still worth pointing out that the results in this work are also applicable to other 
internal collision relevant GRB models, such as the Internal- Collision-induced MAgnetic Reconnection and Turbulence (ICMART) model, which effectively increase the $f$ value by invoking a more efficient radiation process \citep{zhangyan11}.  

3) When defining effective internal collisions for $\g$-rays, the results is dependent on the adopted detector sensitivity limit, since sometimes the late time internal collisions only give marginally detectable signals. We justify this effect by decreasing $L_{\rm det,\g}$  to $10^{-9}~ \rm{erg~s^{-1}~cm^{-2}}$ and re-simulate all the results for $T_{\rm pr,obs}$ (green lines in Figures 1). It turns out that the distribution of  $T_{\rm pr,obs}$ becomes slightly more spread out for lower detector sensitivities, but does not affect our main conclusions above. Note that in a  more precise treatment, the detailed detector trigger function would also be required \citep{levan14}.

4) When defining effective internal collisions for X-rays, we used an X-ray detector sensitivity limit as a threshold. In reality, one needs to compare the peak flux of the relevant X-ray flares with the baseline of X-ray afterglows from the external shock. We did not calculate the external shock afterglow lightcurves since too many additional unknown free parameters would be involved. This might only affect the results concerning $T_{\rm ce,obs}$ for very limited cases, i.e., one-episode injection cases and $T_{\rm ce,obs}$ are determined by weak residual internal collisions rather than refreshed collisions.

5) In the simulations, we assume a constant density for the circumburst medium, with 
$n=1~\rm cm^{-3}$. If the actual density baseline is different, the dynamics of the external blastwave and thus its confinement of the late internal collisions would be affected, although the 
dependence of the blastwave dynamics on the value of $n$ is relatively small, as shown from 
equations \ref{eq:rdec} and \ref{BM}. We have tested such effects through some additional simulation 
runs for four constant values of the circumburst number density, e.g., $n=0.1~\rm cm^{-3}$, 
$n=1~\rm cm^{-3}$, $n=10~\rm cm^{-3}$ and $n=100~\rm cm^{-3}$. It turns out that for specific models 
such as the plausible models 1, 7 and 9, the deviation among the results for the different number 
densities are less than a factor of 2 for both $T_{\rm pr,obs}/T_{\rm ej}$ and $T_{\rm ce,obs}/T_{\rm ej}$. 
On the other hand, the medium density could be of a wind type instead of being constant. If so, the blastwave tends to start deceleration at a lower radius, which might lead to a stronger confinement of the late internal collisions. The consequences for our results of having such a wind can be estimated as
follows. Since the blastwave confinement starts to work at the deceleration radius, bursts with similar 
values of $R_{\rm dec}$ and $n(R_{\rm dec})$ (the number density at $R_{\rm dec}$) should give similar 
results in our simulations, regardless of the exact density profile. For the commonly used free 
stratified wind model $n=AR^{-2}$, with $A=\dot{M}/4\pi m_pv_w=3\times10^{35}A_*\rm{cm^{-1}}$ and
$A_*=(\dot{M}/10^{-5}~\rm{M_{\odot}~yr^{-1}})(v_w/10^3\rm{~km~s^{-1}})^{-1}$ \citep{chevalier99,chevalier00}, the deceleration radius is $R_{\rm dec}=E/(2\pi Am_pc^2\g_0^2)$. Comparing with our equ.(\ref{eq:rdec}), we have $R_{\rm dec}^{\rm wind}/R_{\rm dec}^{\rm const}=1.1E_{53}^{2/3}n_2^{1/3}A_{*,-1}^{-1}\g_{0,2}^{-4/3}$, and $n(R_{\rm dec}^{\rm wind})/n(R_{\rm dec}^{\rm const})=0.25E_{53}^{-2}n_2^{-1}A_{*,-1}^{3}\g_{0,2}^{4}$. The results depend sensitively on the value of $A_*$, which is very uncertain due to our poor 
knowledge of the GRB progenitors, and in principle it could vary from 0.01 to 10 \citep{chevalier99}. 
One sees that for $A_*<1$ cases, as far as the confinement effect, the wind type medium should be 
comparable to a constant medium with $n\sim10^{2} ~\rm{cm^{-3}}$, with the inference that for a wind 
medium, our results might be altered a factor of $\sim$ 2. Although $A_\ast \lesssim 1$ values are
plausible, it is worth pointing out that more severe deviations might occurr for a wind type medium 
if $A_*$ is much larger than unity. 

6) Within one ejection episode, the initial Lorentz factors of all shells, no matter whether
earlier or later ejected, are chosen randomly out of certain distribution functions (RAN, PL or 
GAUSS). The temporal variability of the distribution is not considered, since this is related to the 
activity details of the GRB central engine, which is very uncertain, and at this point no numerical 
simulations have tackled this difficult problem yet. However, it is worth noting that if 
the distribution functions indeed suffer a temporal evolution, the present results could be altered. 
For instance, if the shells ejected later  tend to have lower initial Lorentz factors, internal 
collisions would become inefficient or even nonoperational at the late stages, leading to smaller 
values of $T_{\rm pr,obs}$. In this case, the value of $T_{\rm ce,obs}$ could become either smaller 
or larger, depending on whether the late ejected shells are energetic enough to trigger effective 
refreshed collisions.

7) When using the results of $t_{\rm burst}$ to estimate $T_{\rm ce}$, a more careful treatment for the selection effects would give better results. Two types of selection effects are essential: 1) effects due to the satellite properties \citep{zhang14}; 2) effects due to the sample selection criteria defined in \cite{zhang14}.

In summary, we find that late internal collisions or refreshed external collisions from early ejected shells result in values of $T_{\rm 90}$ and $t_{\rm burst}$ larger than $T_{\rm ce}$, usually by factors of $2-3$. However this is only valid when $T_{\rm ce}\lesssim 10^4$ s, owing to the confinement by the external blastwave. For $T_{\rm ce}\gtrsim 10^4$ s cases, $t_{\rm burst}$ is always a good estimator for $T_{\rm ce}$, and $T_{90}$ might be much smaller than $T_{\rm ce}$ when the late central engine activity is moderate. These conclusions are insensitive to the initial condition assumed for the models.
We have used the results of our simulations to estimate the intrinsic distribution of central engine activity times $T_{\rm ce}$, making use of the observational data for $T_{90}$ and $t_{\rm burst}$. The results suggest a clear bimodal distribution of $T_{\rm ce}$, a conclusion which is insensitive to the initial conditions used in the  models. Based on these results, we conclude  (i) that bursts with $T_{90}$ of order of $10^3$ s need not belong to a special population, and (ii) that their intrinsic central engine activity time could be substantially smaller than 300 s. Bursts with $t_{\rm burst} > 10^4$ s might, on the other hand, belong a new population, and $t_{\rm burst}$ appears to be a reliable measure to define ``ultra-long" GRBs.
Our results are also compatible with the late central engine activity of such bursts tending to become much more moderate than the early activity, as might be expected from a tapering off of the amount of fall-back material onto the central engine.  Further observations and a larger sample of ultra-long GRBs will be required before firmer conclusions about this model can be drawn.

\acknowledgments{We thank Bing Zhang and Bin-Bin Zhang for helpful comments, and an anonymous referee for a constructive report. This research was supported in part by NASA NNX 13AH50G.}

\clearpage

\begin{table}[H]
\label{table:time}
 \begin{center}{\scriptsize   
\begin{tabular}{ll}\hline\hline
Symbol & ~~~~~~~~~~~~~~~~~~~~~~~~~~~~~~~~~~~~~~~~Definition\\ \hline
$T_{\rm ce}$ & Intrinsic central engine activity timescale\\
$T_{90}$ &  Time interval over which $90\%$ of the total background-subtracted counts are observed in  $\g$-ray band \\
$t_{\rm burst}$ & Burst duration defined based on both $\g$-ray and the X-ray light curve features \citep{zhang14} \\
$T_{\rm ej}$& Ejection timescale in the simulation,  corresponding to $T_{\rm ce}$ for real source\\
$T_{\rm pr,obs}$& Final estimated prompt emission duration from the simulation, corresponding to $T_{90}$ for real source \\
$T_{\rm ce,obs}$ & Observation inferred central engine timescale in the simulation, corresponding to $t_{\rm burst}$ for real source\\
$T_{\rm col,lab}$ & Last mechanical collision time in lab frame during internal shock phase  \\
$T_{\rm eff,lab}$& Last effective collision time in lab frame during internal shock phase\\
$T_{\rm col}$&$T_{\rm col,lab}$ value in observer frame\\
$T_{\rm eff}$ & $T_{\rm eff,lab}$ value  in observer frame\\
$T_{\rm af}$& Timescale for last identifiable internal signature in X-ray afterglow light curve\\
$t_{\rm col}$ & Collision time for two shell case in lab frame \\
$t_{\rm col,obs}$ &$t_{\rm col}$ value in observer frame \\
$\delta t_e$& Emission duration for internal shock in lab frame \\
$\delta t_{\rm obs}$& Emission duration for refreshed shock in observer frame \\
$t_{\rm lab,0}$ & Initial lab frame time in one simulation  \\
$t_{\rm lab,n}$ & Lab frame time for the $n$th collision\\
$t_{\rm next}^n$ & Time interval between ($n-1$)th and $n$th collision in lab frame \\
$t_n$& Detection time of $n$th collision by observer at luminosity distance $D_L$ in lab frame \\
$t_{\rm obs,n}$ & Timescale for $n$th collision since detector trigger time in observer frame   \\
$\delta t_{\rm obs,n}$ & Observer frame emission duration for $n$th collision \\
$t_{\rm obs,n\g}$ & Time for last effective internal collision for $\g$-rays in observer frame \\
$\delta t_{\rm obs,n\g}$& Emission duration for last effective internal collision for $\g$-rays in observer frame \\
$t_{\rm obs,nX}$ & Time for last effective internal collision for X-rays in observer frame\\
$\delta t_{\rm obs,nX}$ &Emission duration for last effective internal collision for X-rays in observer frame\\
$t_{\rm obs,nR}$ & Time for last effective refreshed collision in observer frame \\
$\delta t_{\rm obs,nR}$ & Emission duration for last effective refreshed collision in observer frame\\
$\Delta t_{\rm ej}$ & Ejection time intervals between shells \\
$T_{\rm quie}$& Time intervals between shells injection episodes\\
\hline
\end{tabular}   
}
   \end{center}
  \caption{Notation list for different timescales showing in this work.}
  \end{table} 
   
\clearpage

\begin{table}[H]
\label{table:model}
 \begin{center}{\scriptsize
 \begin{tabular}{c|c|c|c|c|c|c|c} \hline\hline
Model   &  $\g$     &  $\Delta t_{\rm ej}$ (s)         & Episode        & $T_{\rm quie}$ (s) & $\rm log_{10}(m)$ (g) &  $l/c$ (s) &   Figure \\
\hline
 1   & RAN (50,500)       &  RAN (0.5,2.5)            &  1       &  $--$ &  RAN (27,29)     &  RAN (0.1,0.5)            &  $1a,1b$   \\
     \hline
 2   &  RAN (500,1000)      &  RAN (0.5,2.5)           &  1        &  $--$ &  RAN (27,29)     & RAN (0.1,0.5)            & $1c,1d$        \\
  \hline
3     &  RAN (50,500)     &  RAN (5,25)           & 1       &  $--$ &  RAN (27,29)    & RAN (0.1,0.5)  &  $1e,1f$      \\
\hline
4   &  PL (50,500,-1)    &  RAN (0.5,2.5)           &  1       &  $--$ &  RAN (27,29)     &  RAN (0.1,0.5)  &  $1g,1h$       \\
     \hline
5   &  PL (50,500,-0.5)     &  RAN (0.5,2.5)           &  1       &  $--$ &  RAN (27,29)     & RAN (0.1,0.5)  &  $--$       \\
\hline
6   &  PL (50,500,-0.1)     &  RAN (0.5,2.5)           &  1       &  $--$ &  RAN (27,29)     & RAN (0.1,0.5)  &  $--$       \\
\hline
7   &  PL (100,500,-0.5)     &  RAN (0.5,2.5)           &  1       &  $--$ &  RAN (27,29)     & RAN (0.1,0.5)  &  $1i,1j$       \\
\hline
8   &  PL (100,500,-0.5)     &  RAN (5,25)           &  1       &  $--$ &  RAN (27,29)     & RAN (0.1,0.5)  &  $--$       \\
\hline
9   &  GAUSS (300,100)     &  RAN (0.5,2.5)           &  1       &  $--$ &  RAN (27,29)     & RAN (0.1,0.5)  &  $1k,1l$       \\
\hline
10   &  GAUSS (300,100)     &  RAN (5,25)           &  1       &  $--$ &  RAN (27,29)     & RAN (0.1,0.5)  &  $--$       \\
\hline
11     &  RAN (50,500)     &  RAN (0.5,2.5)           & 2       &  $10^3$ &  $\rm RAN (27,29), RAN (27,29)$    & RAN (0.1,0.5)  &  $1m,1n$      \\
\hline
12     &  RAN (50,500)     &  RAN (0.5,2.5)           & 2       &  $10^3$ &  $\rm RAN (27,29), RAN(25,27)$    & RAN (0.1,0.5)  &  $1o,1p$      \\
\hline
13     &  RAN (50,500)     &  RAN (0.5,2.5)           & 2       &  $10^4$ &  $\rm RAN (27,29), RAN (27,29)$    & RAN (0.1,0.5)  &  $--$      \\
\hline
14     &  RAN (50,500)     &  RAN (0.5,2.5)           & 2       &  $10^4$ &  $\rm RAN (27,29), RAN(25,27)$    & RAN (0.1,0.5)  &  $--$      \\
\hline
15     &  PL (100,500,-0.5)     &  RAN (0.5,2.5)           & 2       &  $10^3$ &  $\rm RAN (27,29), RAN (27,29)$    & RAN (0.1,0.5)  &  $--$      \\
\hline
16     &  PL (100,500,-0.5)    &  RAN (0.5,2.5)           & 2       &  $10^3$ &  $\rm RAN (27,29), RAN(25,27)$ & RAN (0.1,0.5)  &  $--$      \\
\hline
17     &  PL (100,500,-0.5)     &  RAN (0.5,2.5)           & 2       &  $10^4$ &  $\rm RAN (27,29), RAN (27,29)$    & RAN (0.1,0.5)  &  $--$      \\
\hline
18     &  PL (100,500,-0.5)    &  RAN (0.5,2.5)           & 2       &  $10^4$ &  $\rm RAN (27,29), RAN(25,27)$ & RAN (0.1,0.5)  &  $--$      \\
\hline
19   &  GAUSS (300,100)     &  RAN (0.5,2.5)           &  2       & $10^3$ &  $\rm RAN (27,29), RAN (27,29)$     & RAN (0.1,0.5)  &  $--$       \\
\hline
20   &  GAUSS (300,100)     &  RAN (0.5,2.5)           &  2       &  $10^3$ &  $\rm RAN (27,29), RAN(25,27)$     & RAN (0.1,0.5)  &  $--$       \\
\hline
21   &  GAUSS (300,100)     &  RAN (0.5,2.5)           &  2       & $10^4$ &  $\rm RAN (27,29), RAN (27,29)$     & RAN (0.1,0.5)  &  $--$       \\
\hline
22   &  GAUSS (300,100)     &  RAN (0.5,2.5)           &  2       &  $10^4$ &  $\rm RAN (27,29), RAN(25,27)$     & RAN (0.1,0.5)  &  $--$       \\
   \hline\hline
 \end{tabular}
 }
 \end{center}
 \caption{Collection of different initial conditions for the simulation and figure numbers corresponding to different models. For one episode cases, $T_{\rm ej}$ is in order of 100 s (1000 s) when $\Delta t_{\rm ej}$ is distributed as RAN(0.5, 2.5) (RAN(5,25)). For two episode cases, $T_{\rm ej}$ is essentially determined by $T_{\rm quie}$.}
 \end{table}

\clearpage

\begin{figure}
\label{fig:simulationI}
    \subfigure[]{
    \includegraphics[width=1.5in]{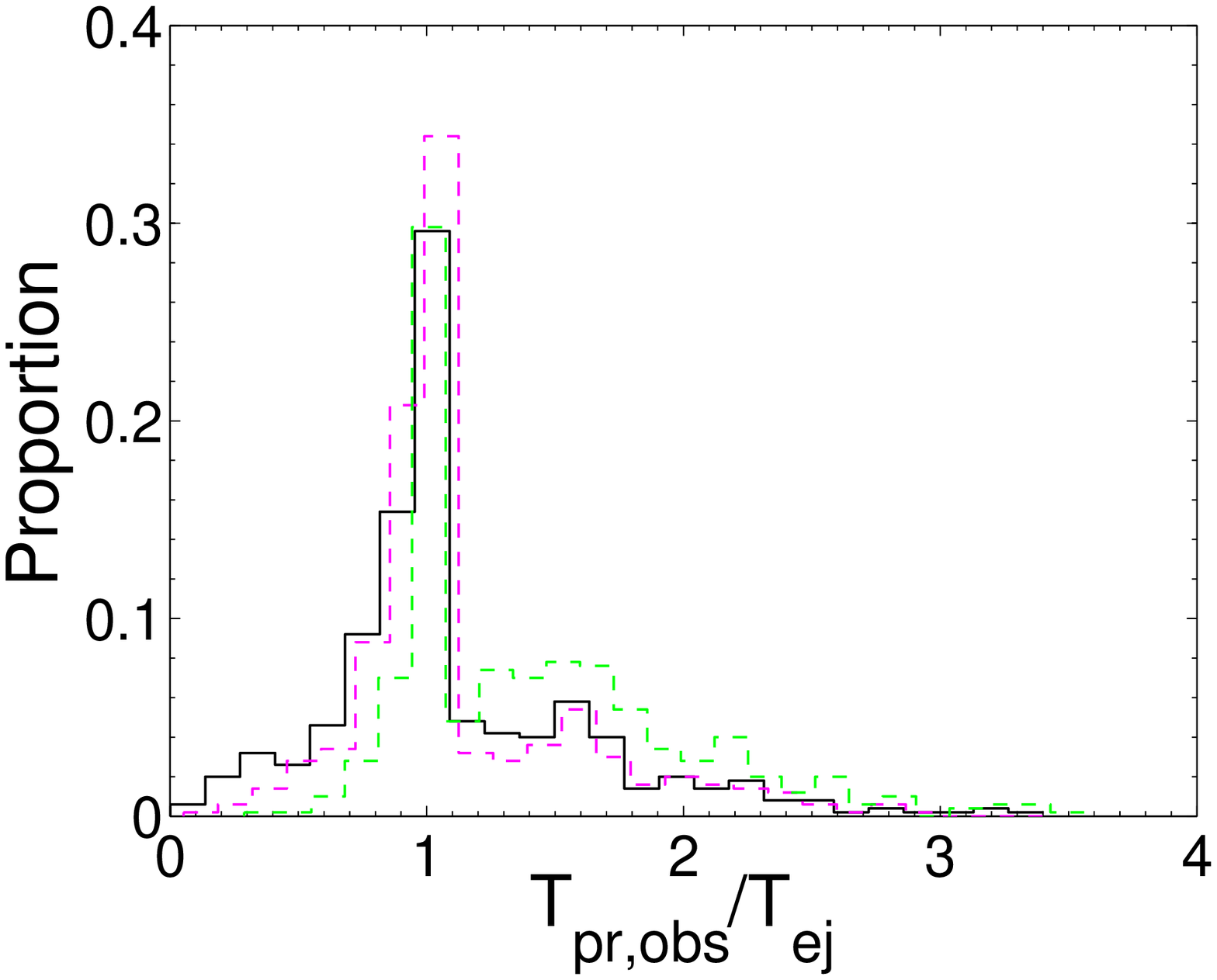}}
    \subfigure[]{
    \includegraphics[width=1.5in]{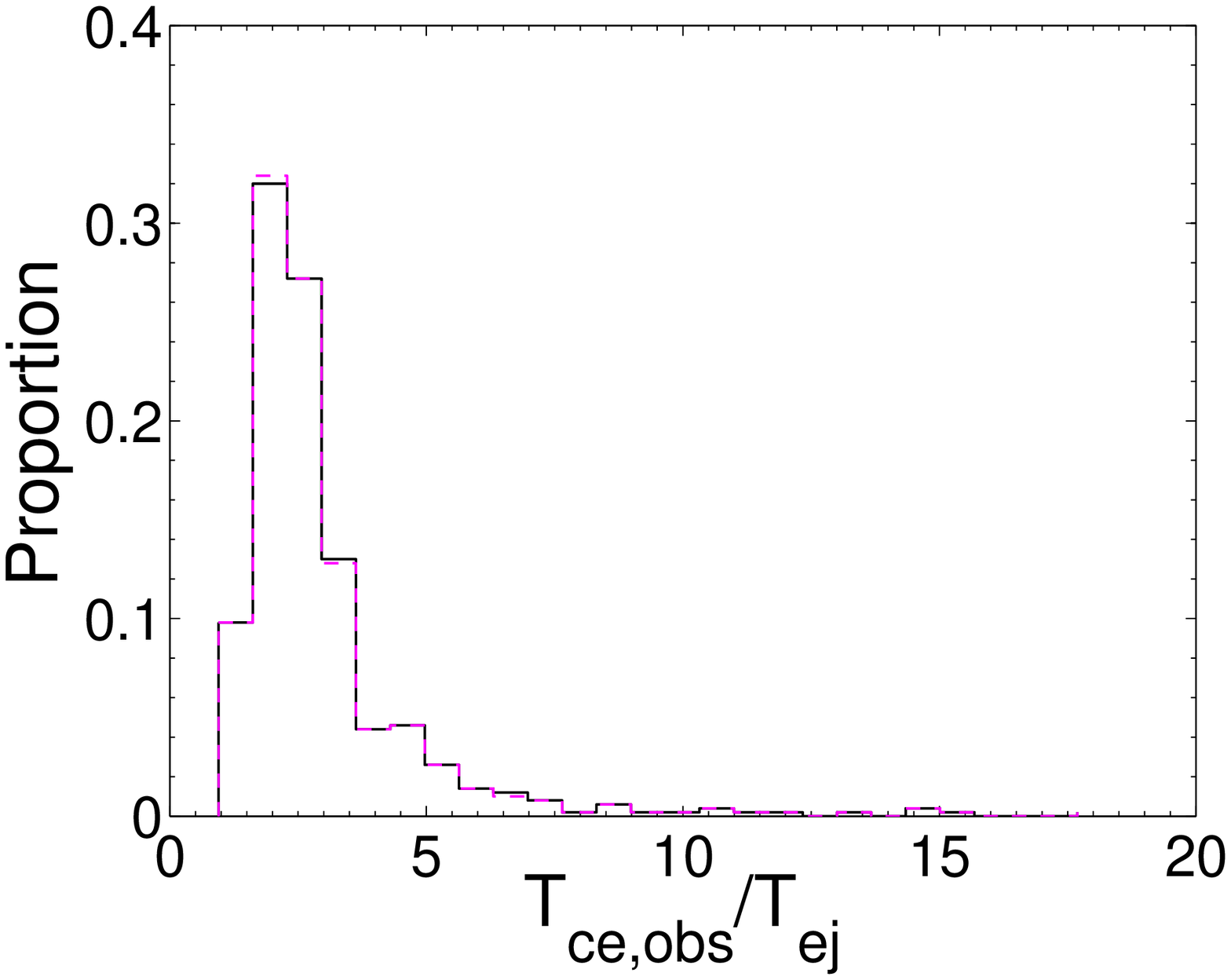}}
     \subfigure[]{
       \includegraphics[width=1.5in]{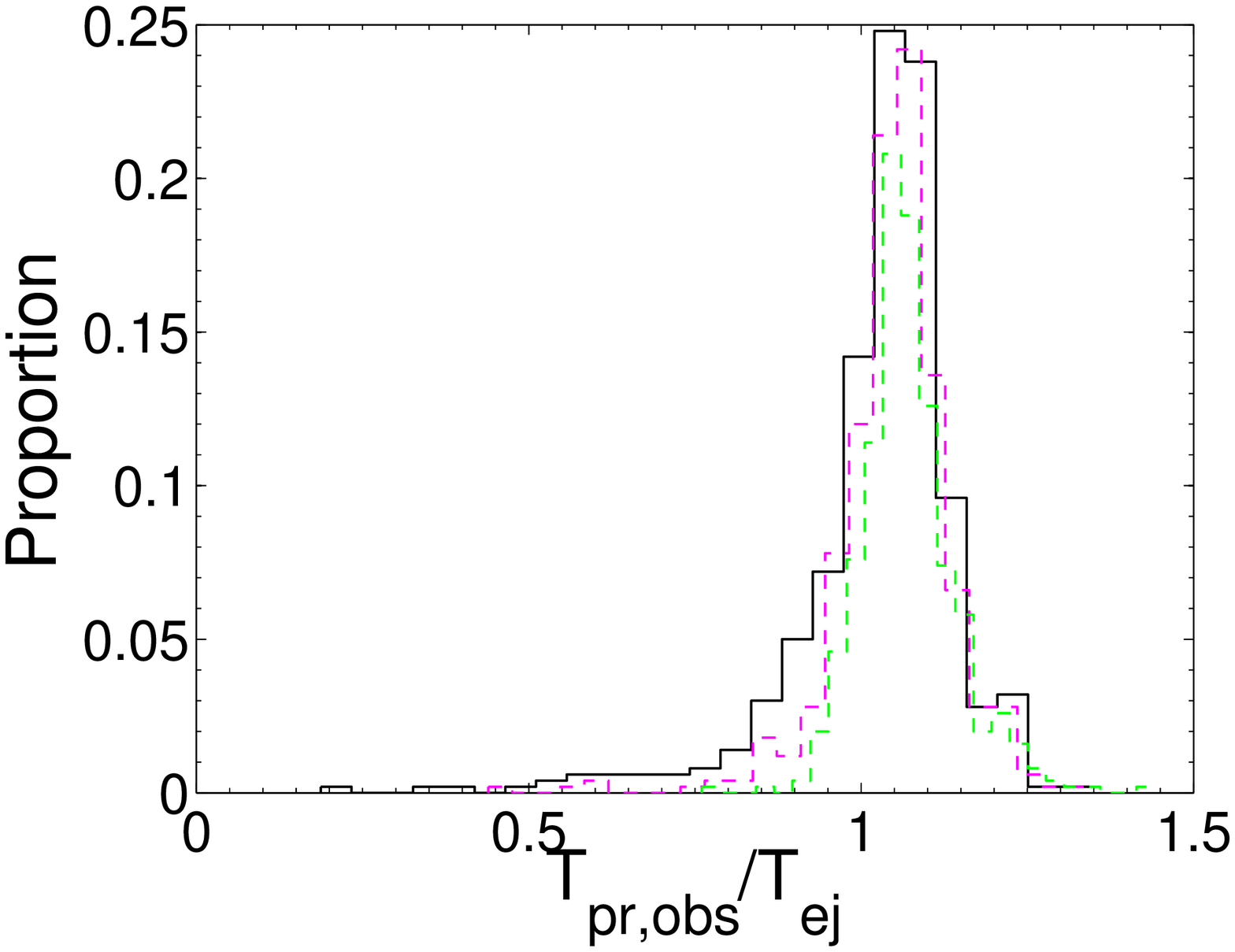}}
    \subfigure[]{
    \includegraphics[width=1.5in]{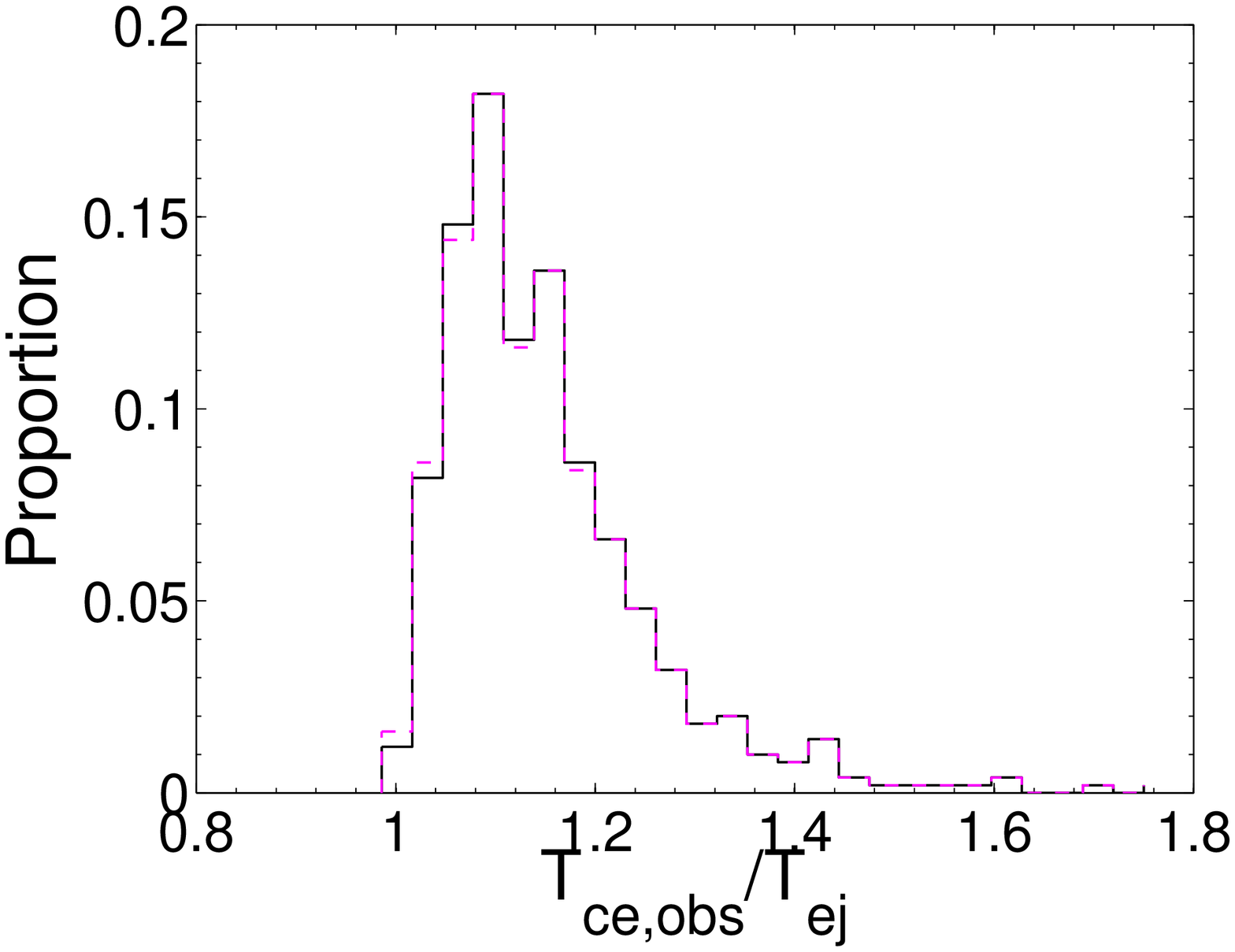}}\\
     \subfigure[]{
    \includegraphics[width=1.5in]{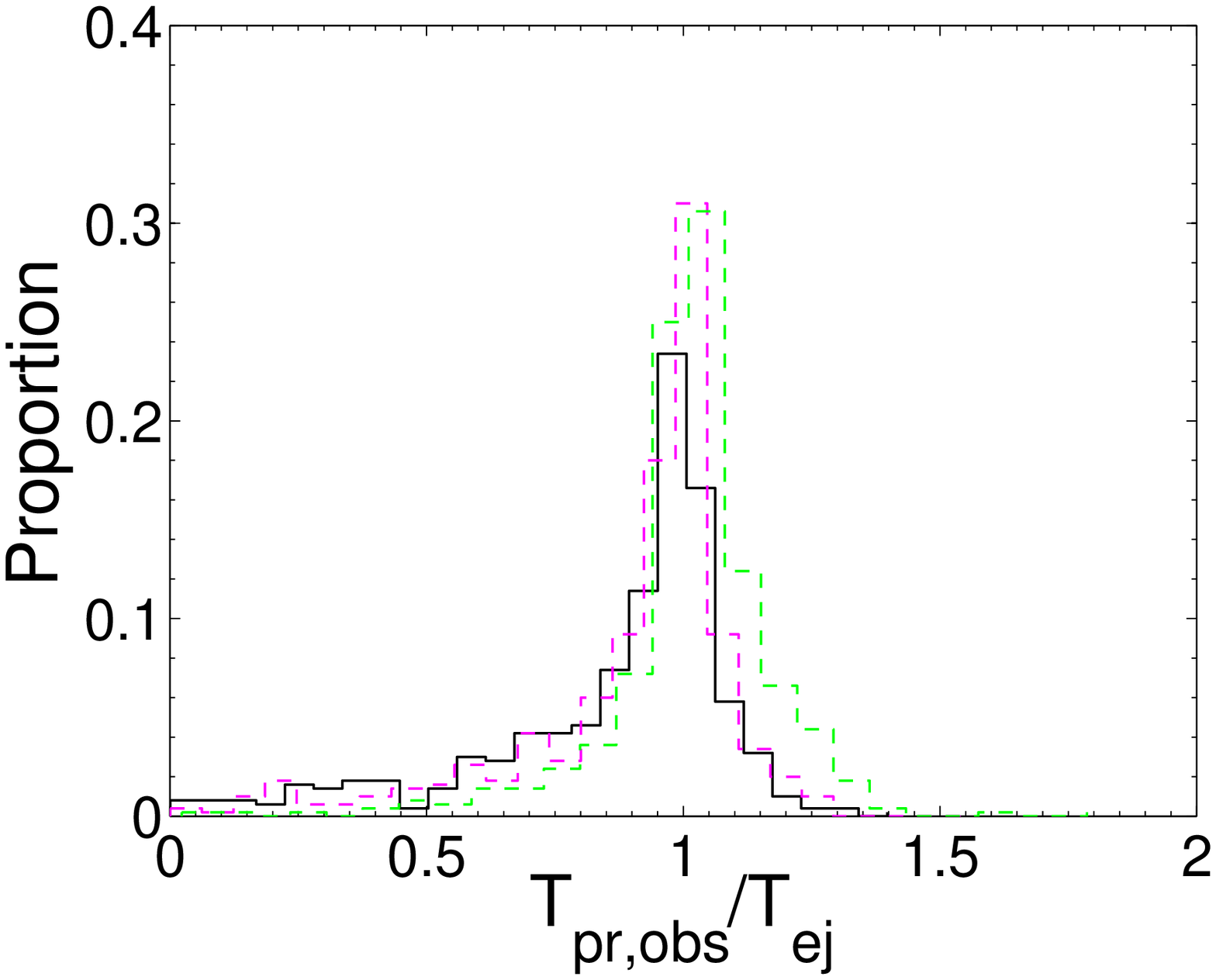}}
    \subfigure[]{
    \includegraphics[width=1.5in]{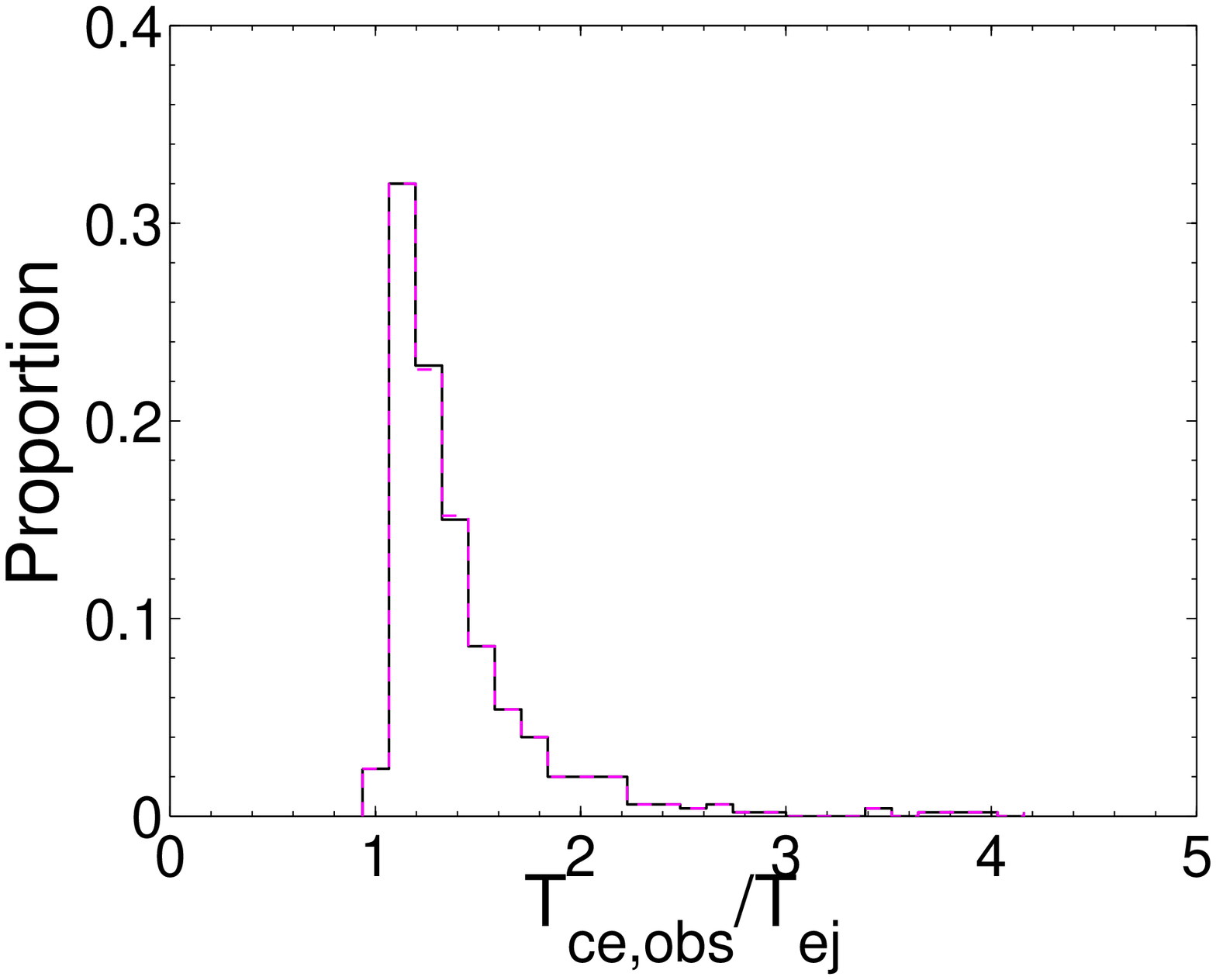}}
     \subfigure[]{
       \includegraphics[width=1.5in]{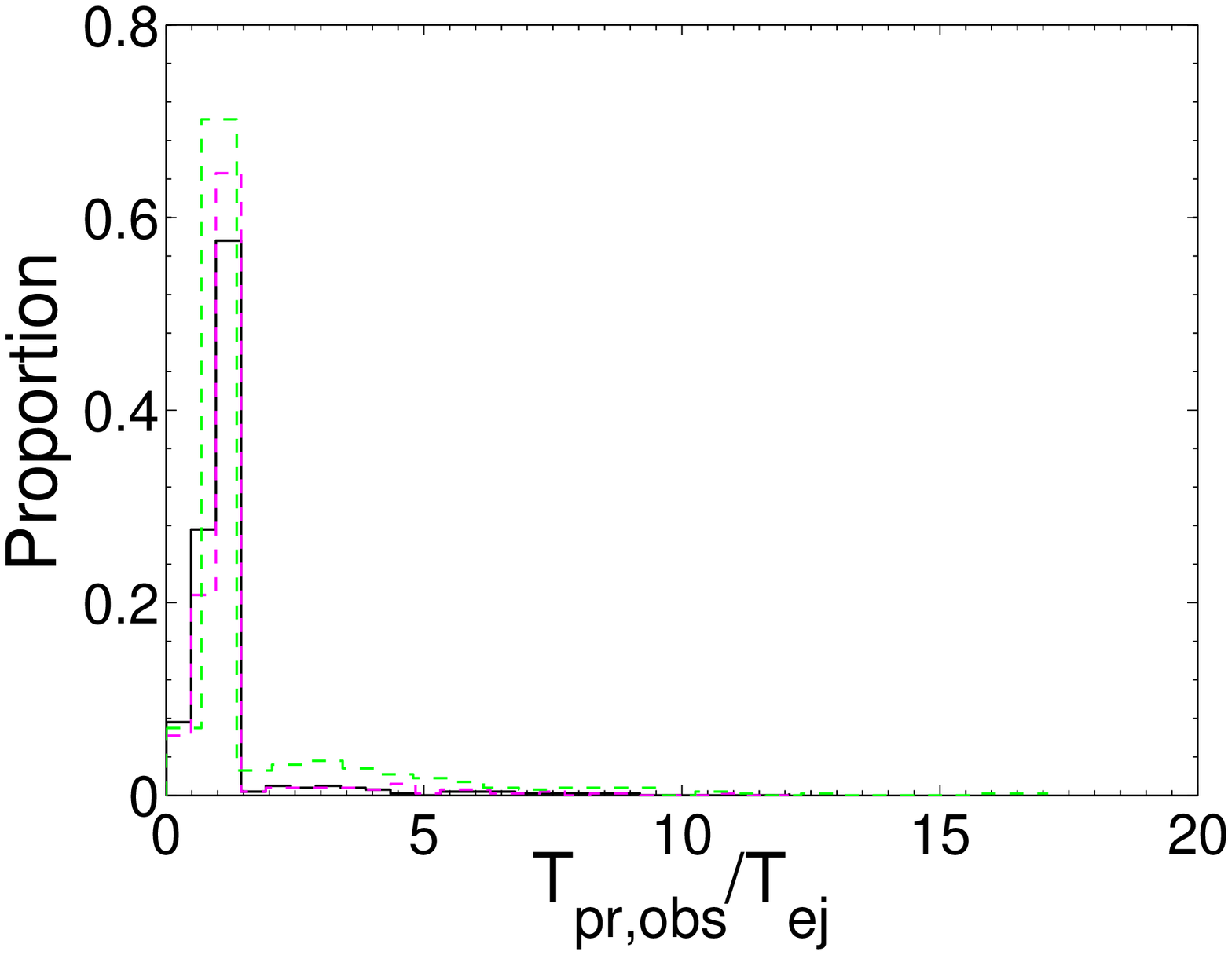}}
    \subfigure[]{
    \includegraphics[width=1.5in]{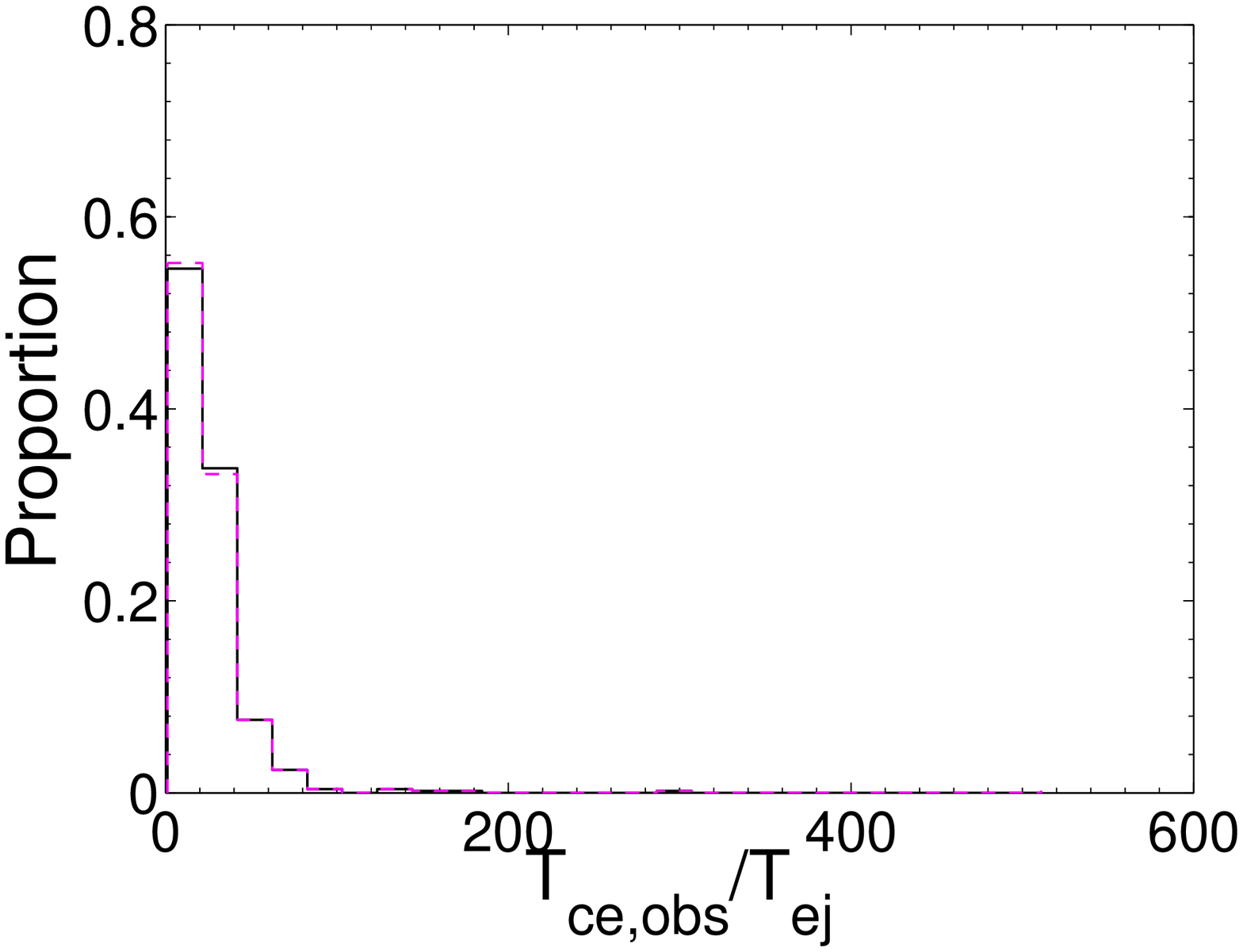}}\\
     \subfigure[]{
    \includegraphics[width=1.5in]{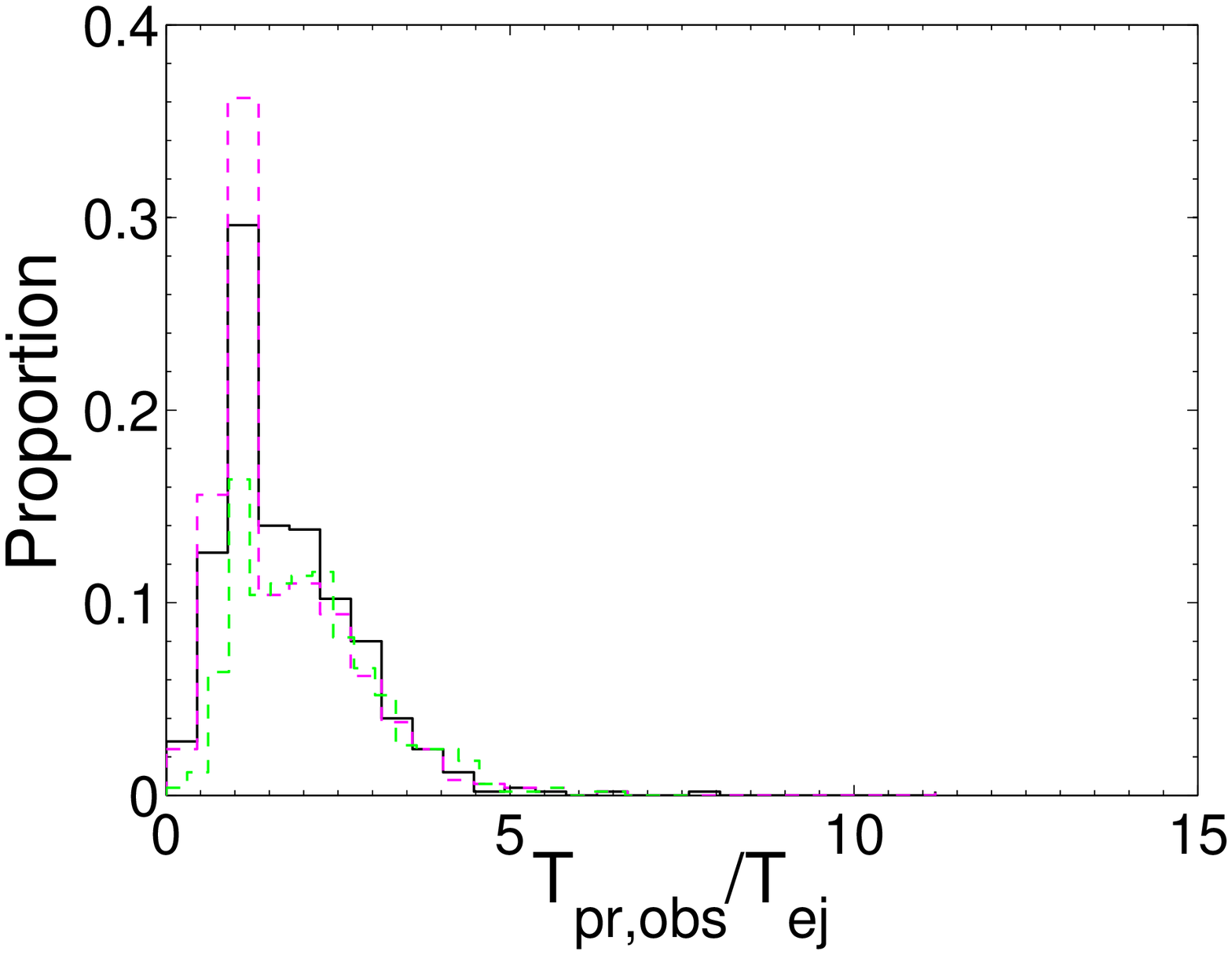}}
    \subfigure[]{
    \includegraphics[width=1.5in]{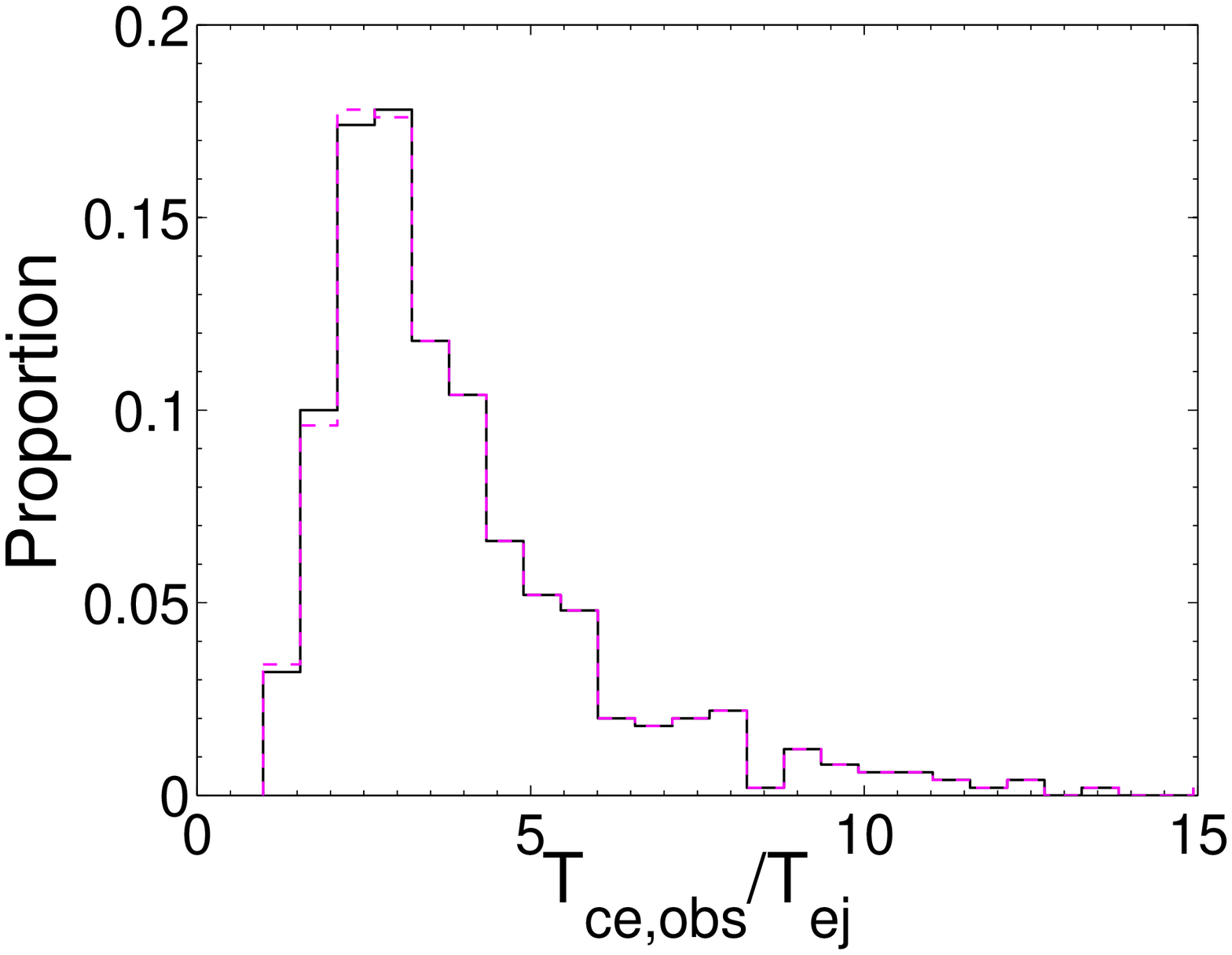}}
     \subfigure[]{
       \includegraphics[width=1.5in]{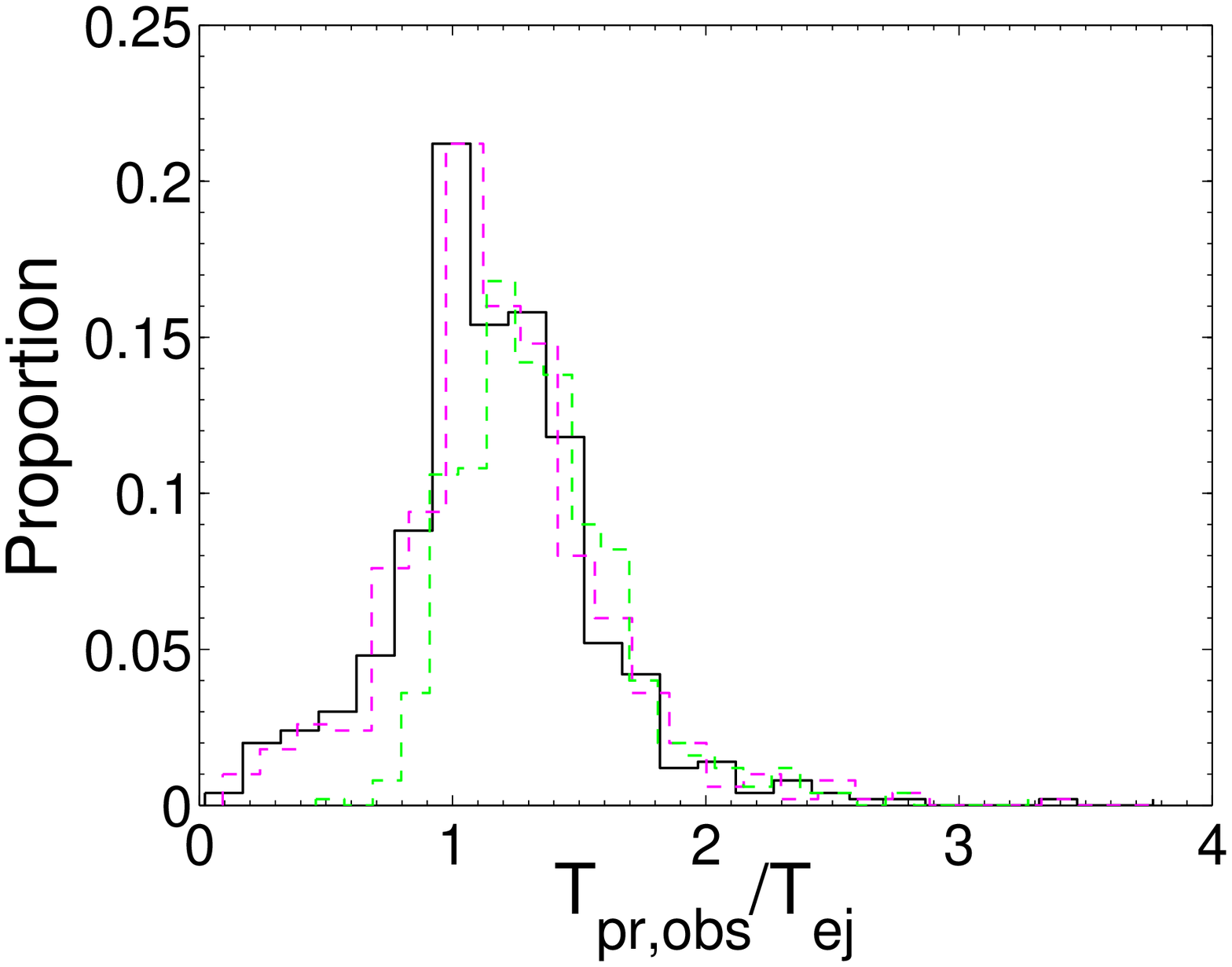}}
    \subfigure[]{
    \includegraphics[width=1.5in]{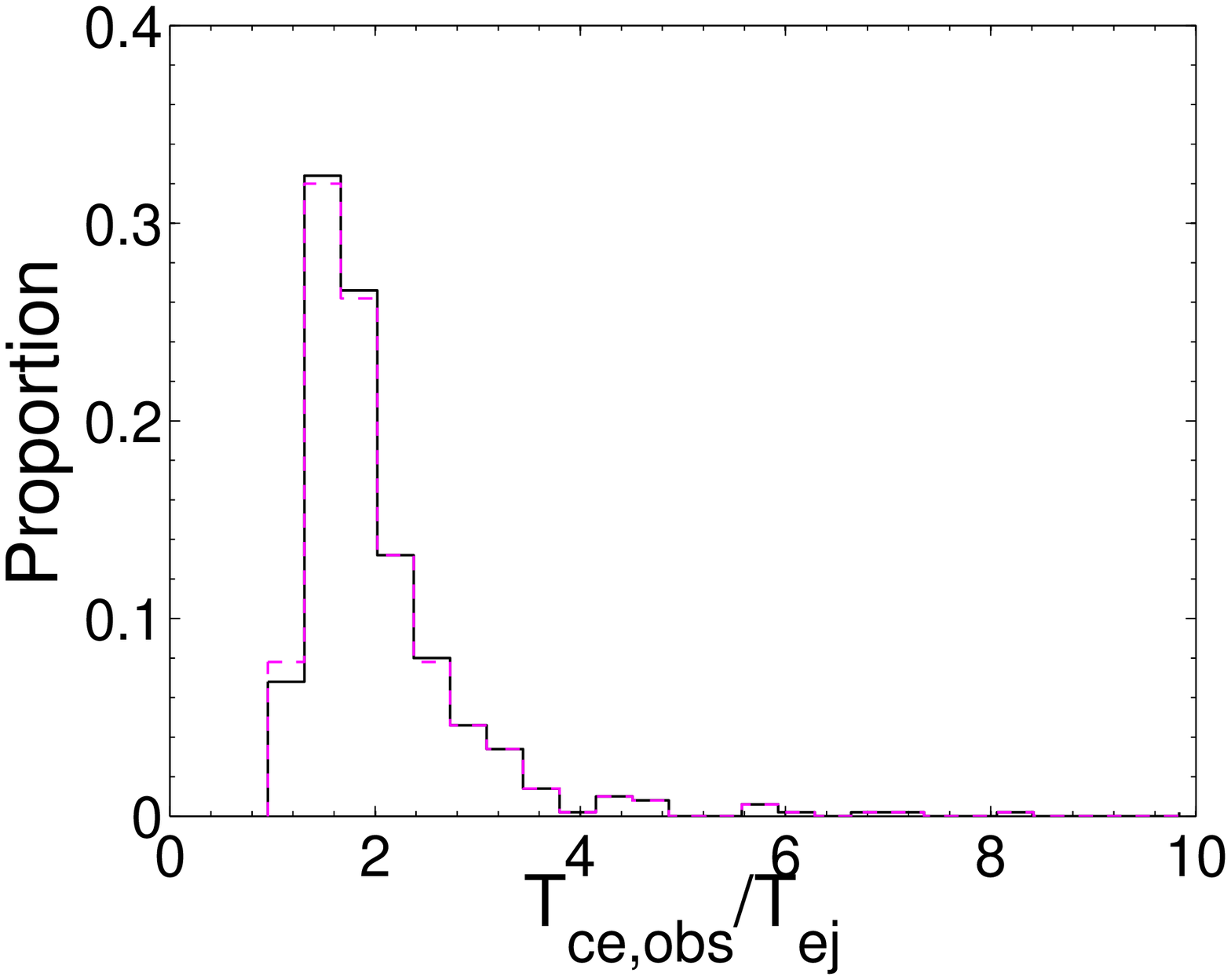}}\\
        \subfigure[]{
    \includegraphics[width=1.5in]{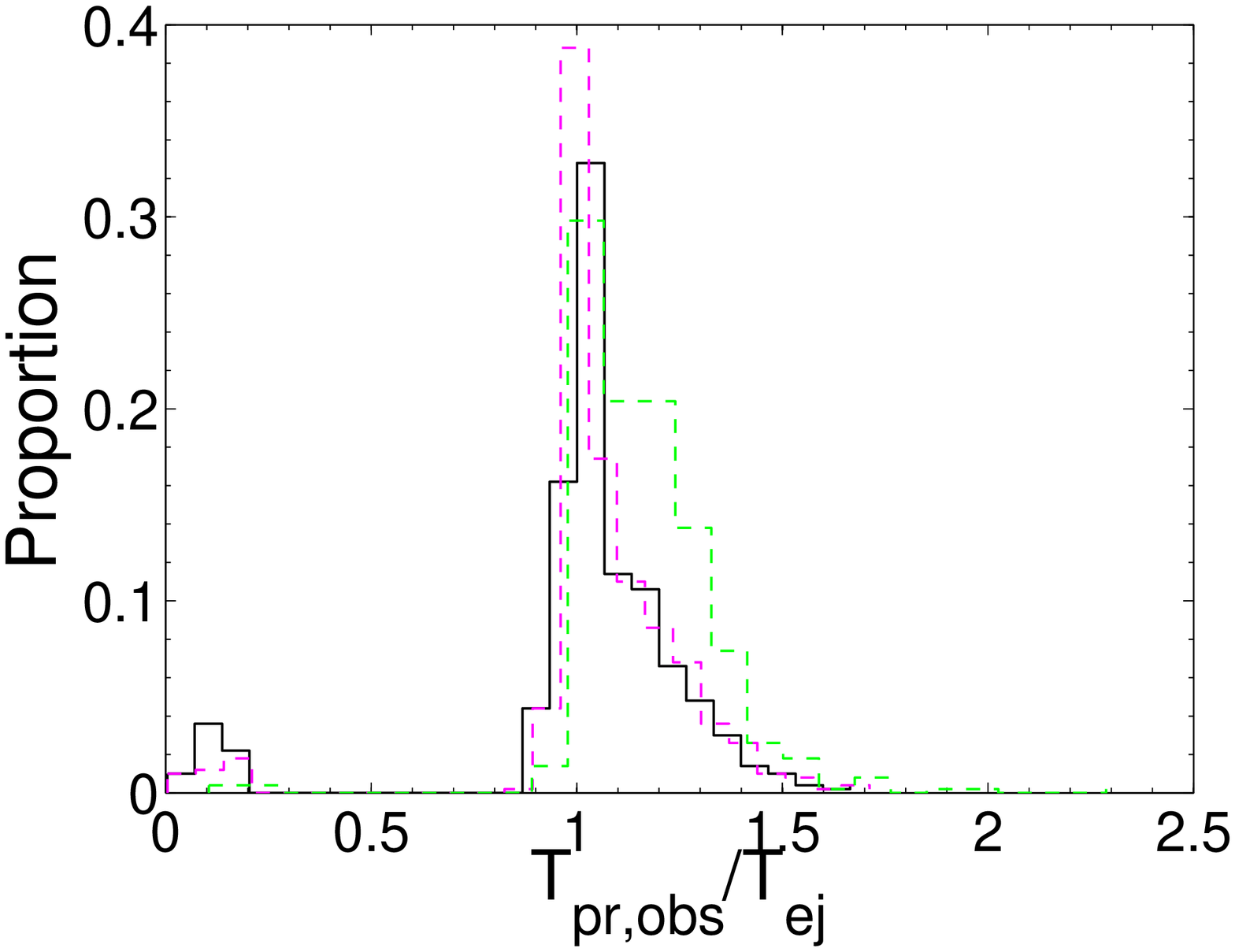}}
    \subfigure[]{
    \includegraphics[width=1.5in]{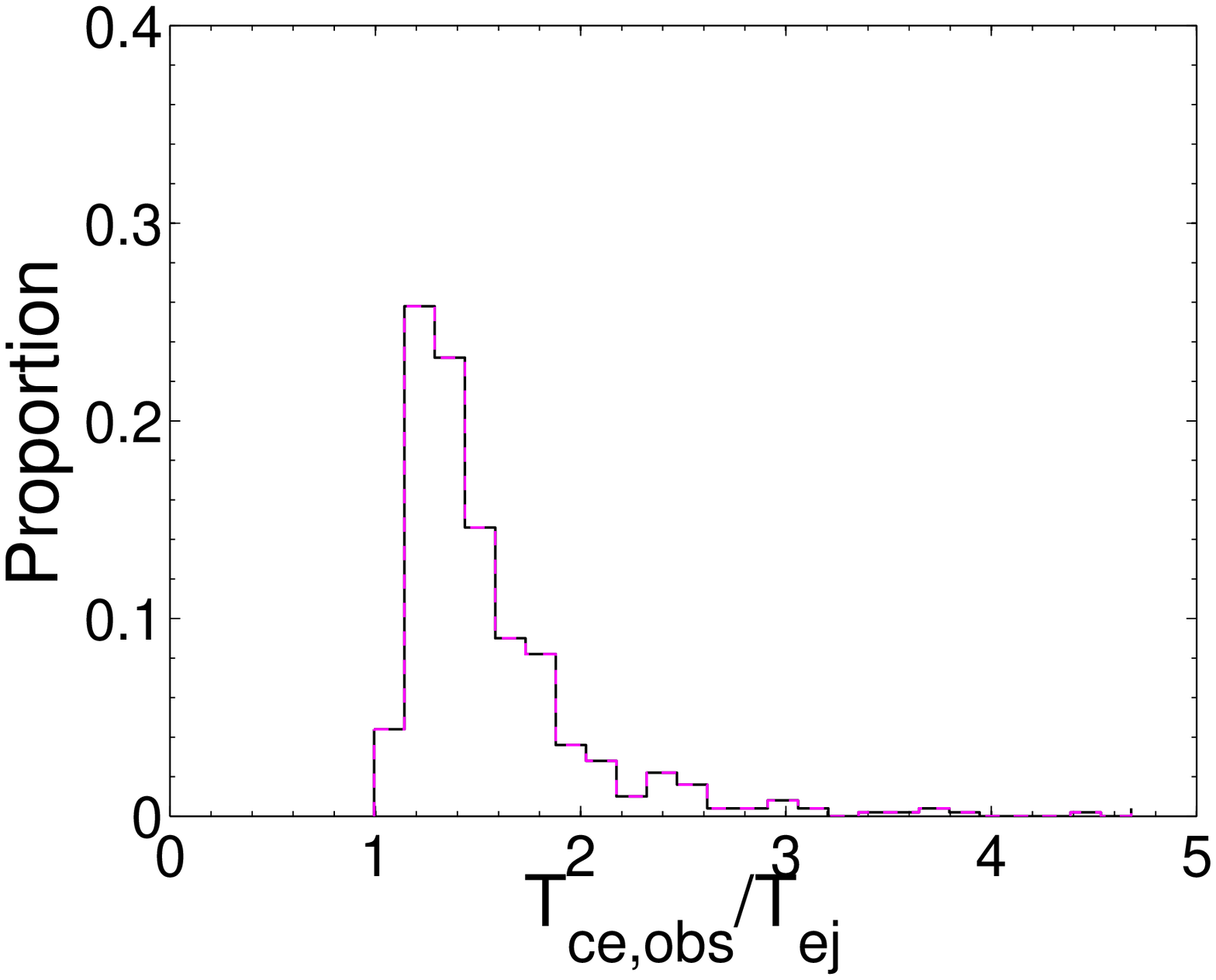}}
     \subfigure[]{
       \includegraphics[width=1.5in]{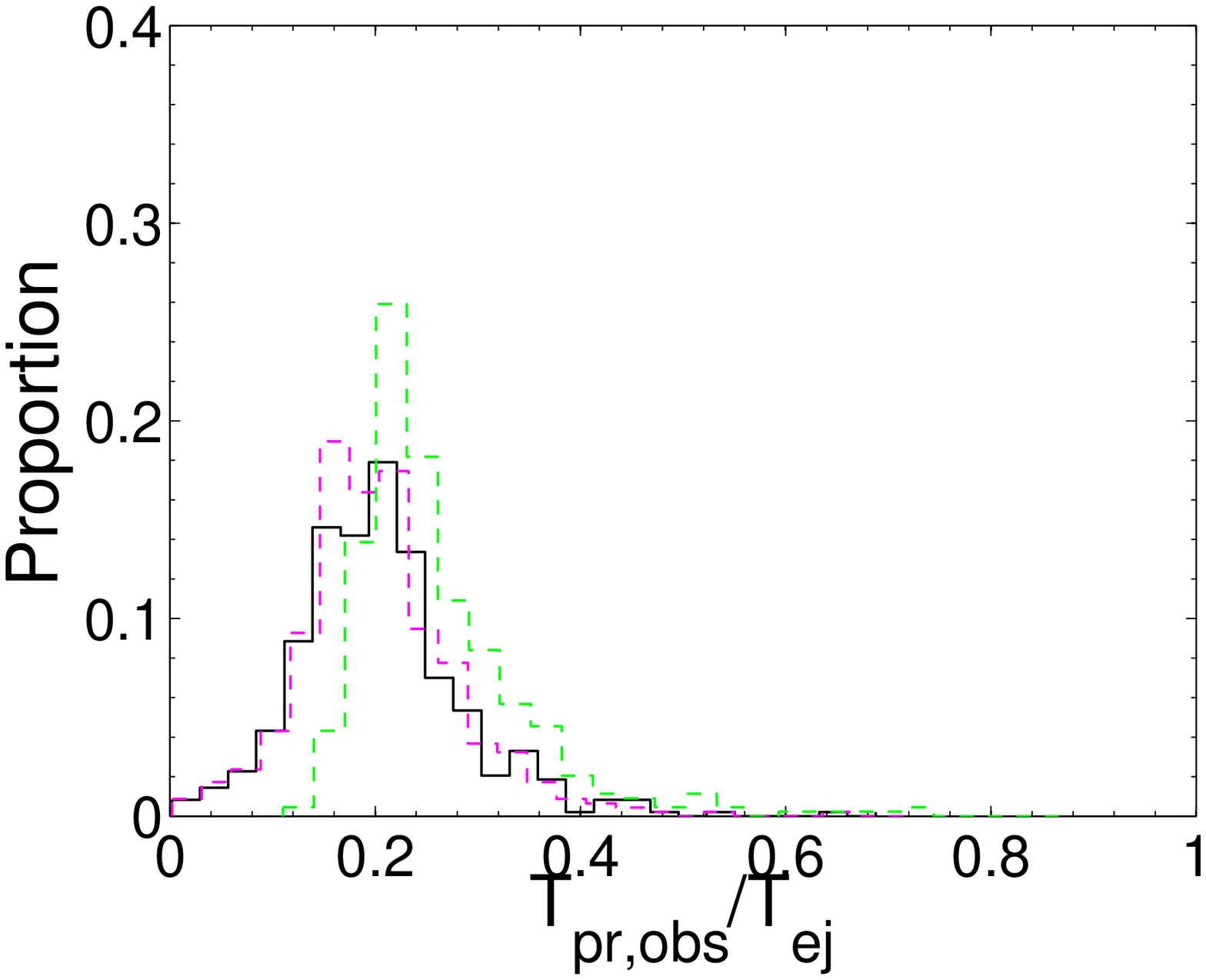}}
    \subfigure[]{
    \includegraphics[width=1.5in]{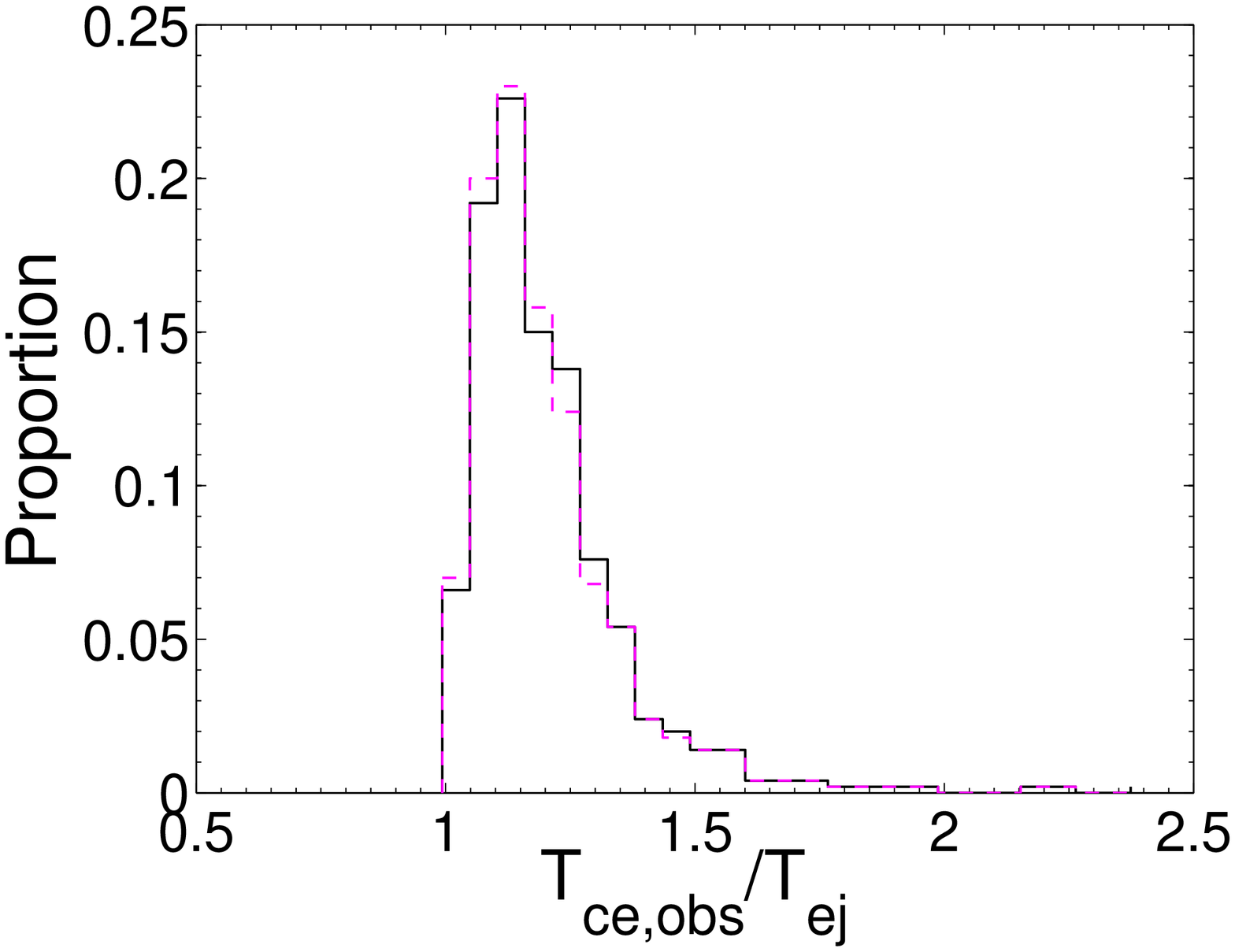}}\\
      \caption{Distributions of $T_{\rm pr,obs}/T_{\rm ej}$ and $T_{\rm ce,obs}/T_{\rm ej}$ for selected models. The corresponding models for each subfigure are summarized in Table 2. Black lines are for $L_{\rm det,\g}=10^{-8}~ \rm{erg~s^{-1}~cm^{-2}}$ and $f=0.5$; pink lines are for $L_{\rm det,\g}=10^{-8}~ \rm{erg~s^{-1}~cm^{-2}}$ and $f=1$; green lines are for $L_{\rm det,\g}=10^{-9}~ \rm{erg~s^{-1}~cm^{-2}}$ and $f=1$.}
            \end{figure}
            
 \clearpage
\begin{figure}
\label{fig:final}
\centering
\subfigure[]{
    \includegraphics[width=2.0in]{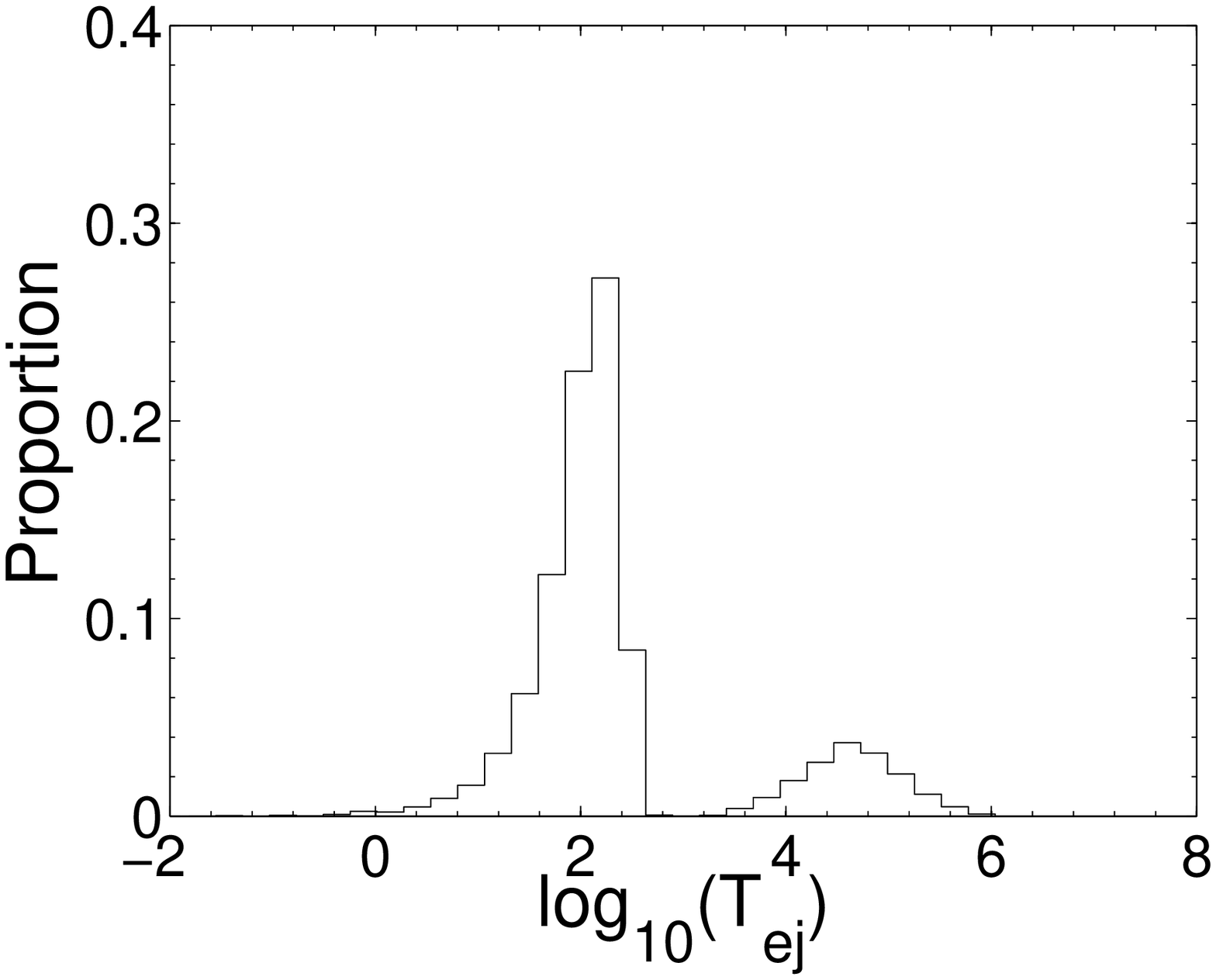}}
    \subfigure[]{
    \includegraphics[width=2.0in]{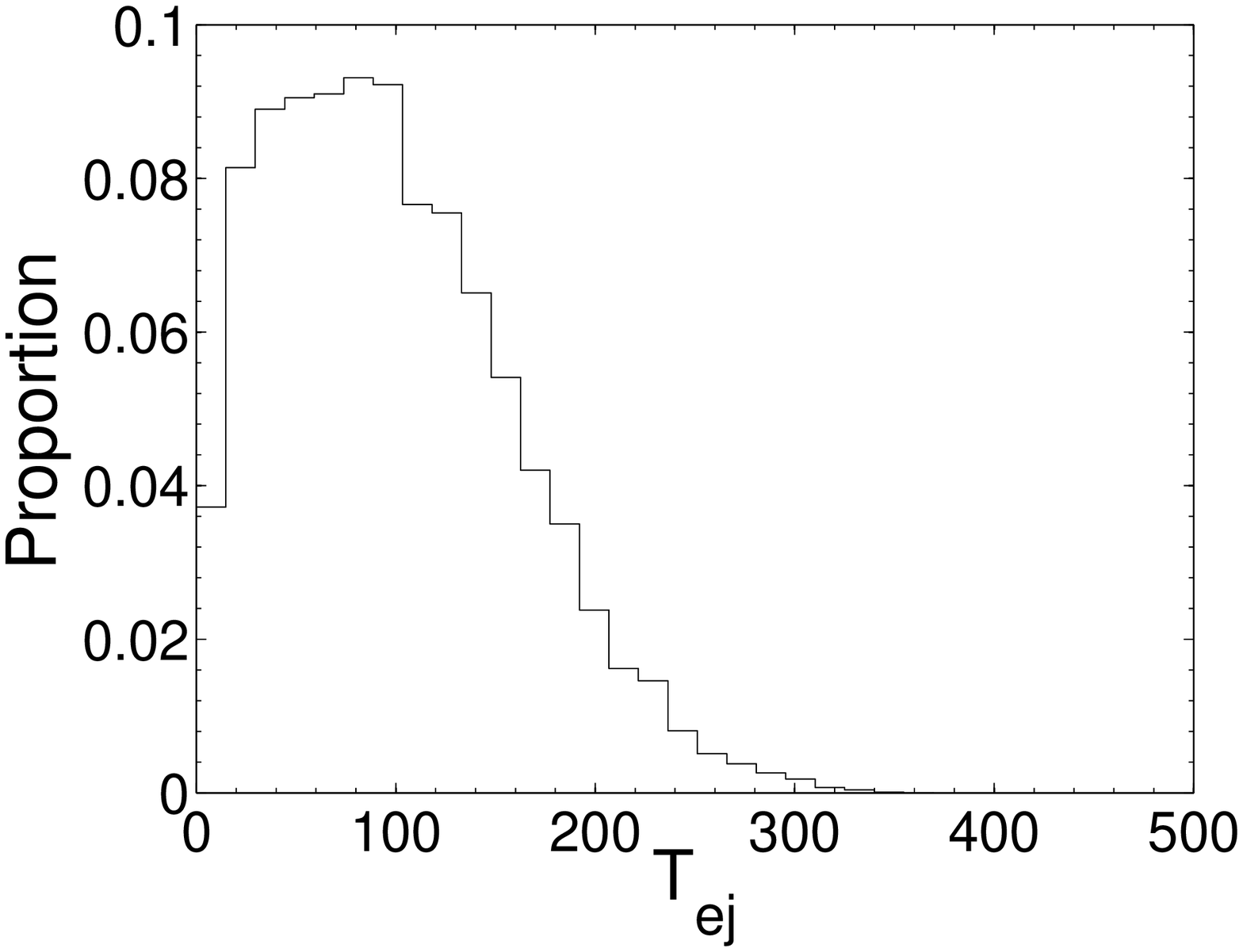}}\\
    \subfigure[]{
    \includegraphics[width=2.0in]{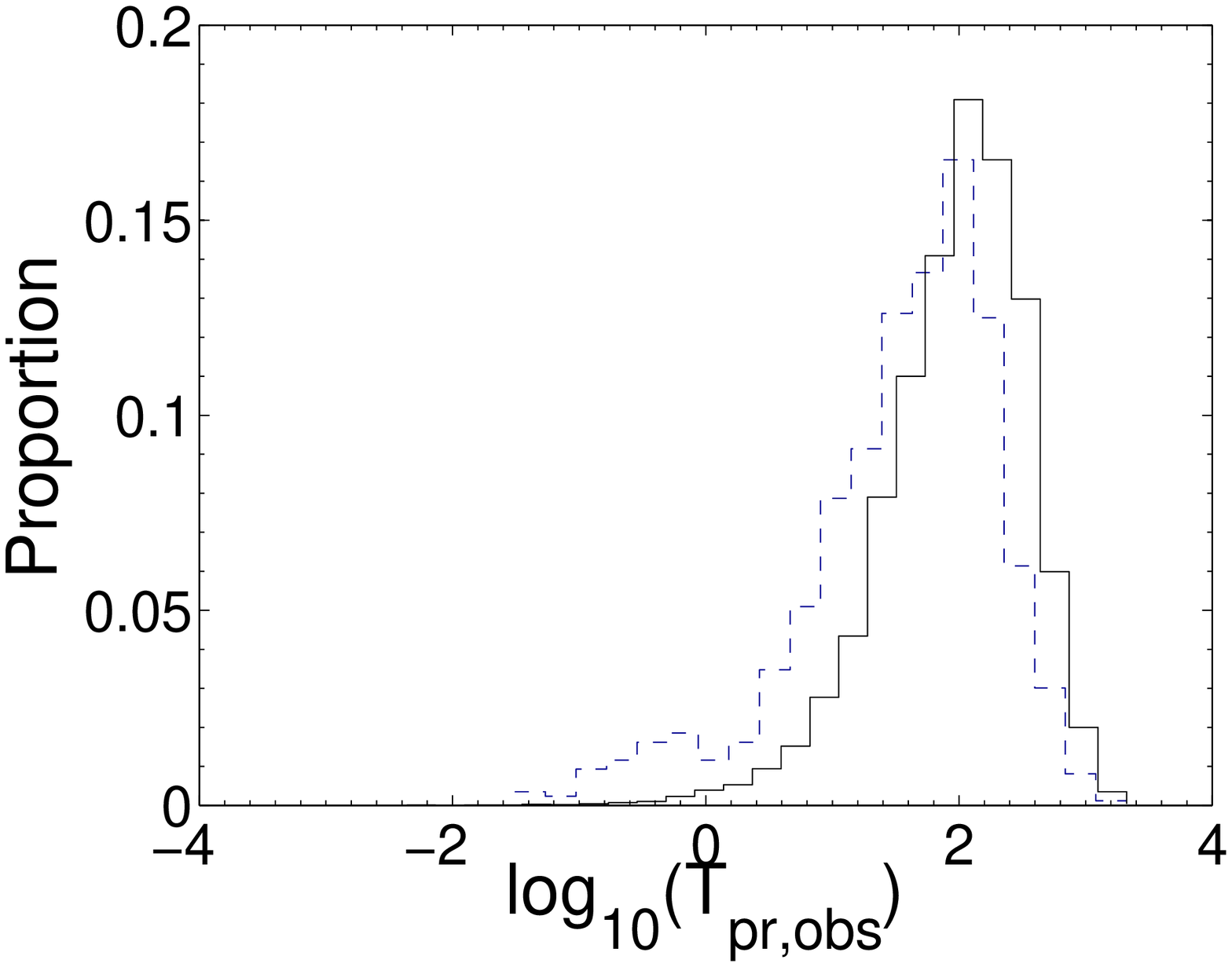}}
    \subfigure[]{
    \includegraphics[width=2.0in]{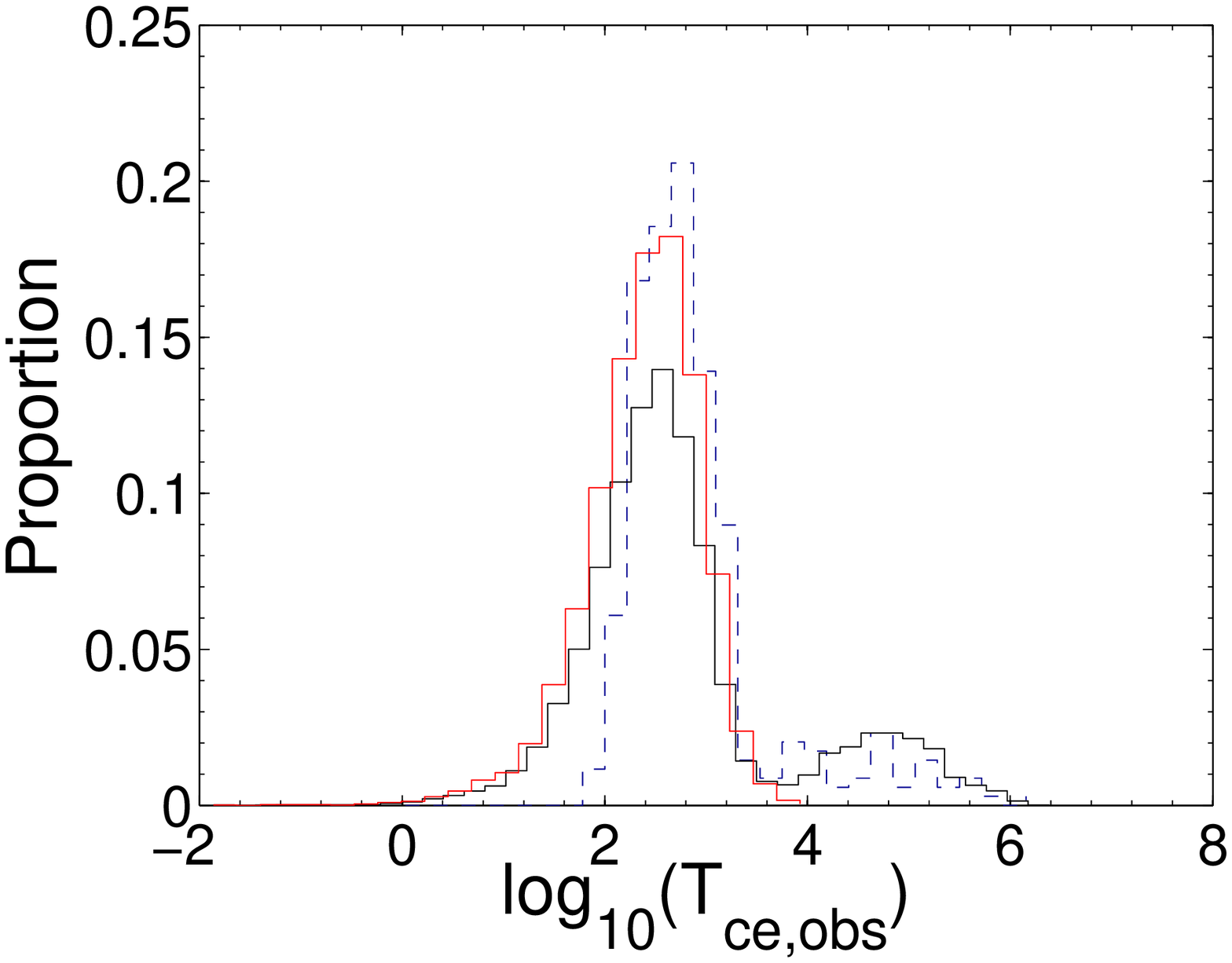}}\\
    \subfigure[]{
    \includegraphics[width=2.0in]{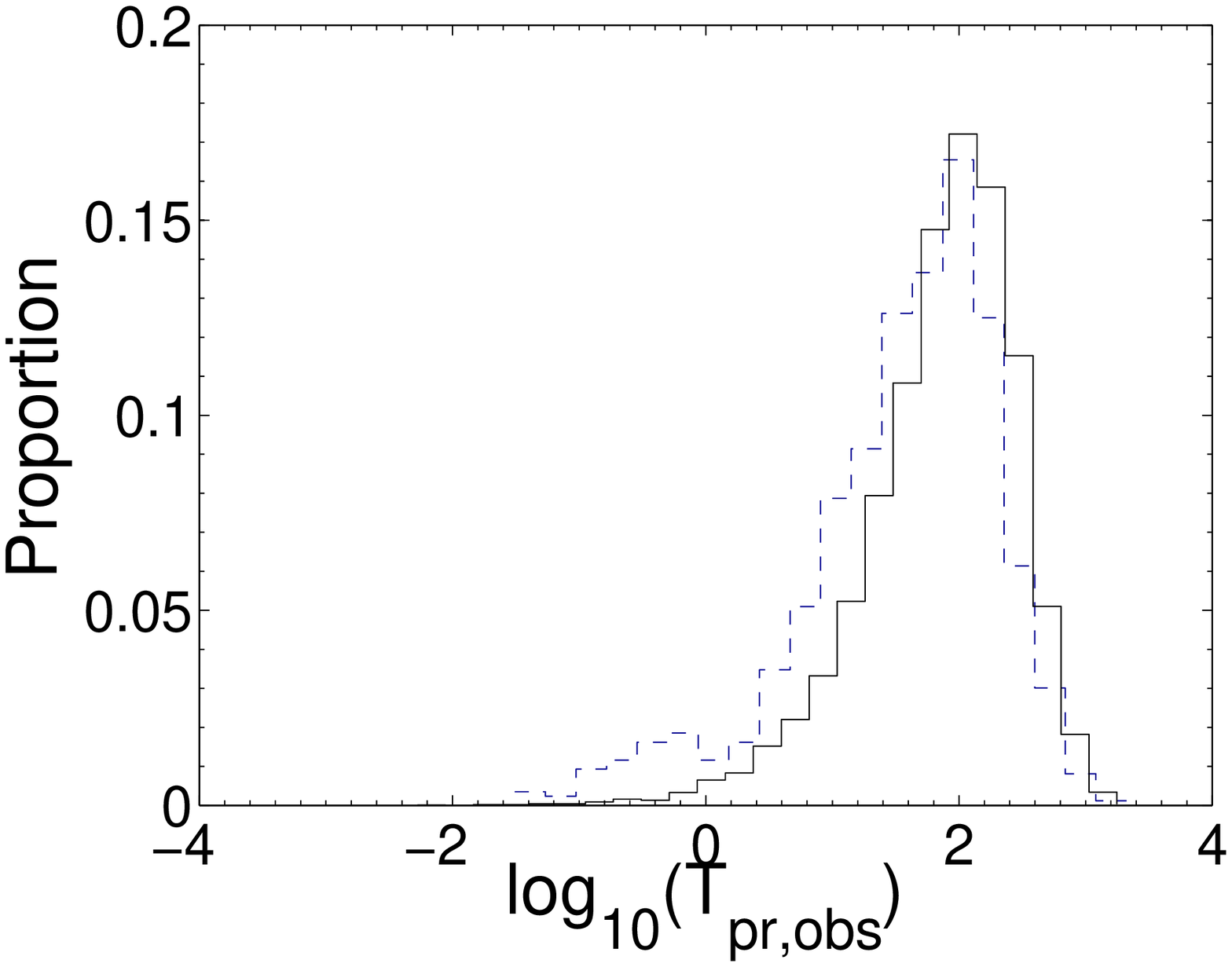}}
    \subfigure[]{
    \includegraphics[width=2.0in]{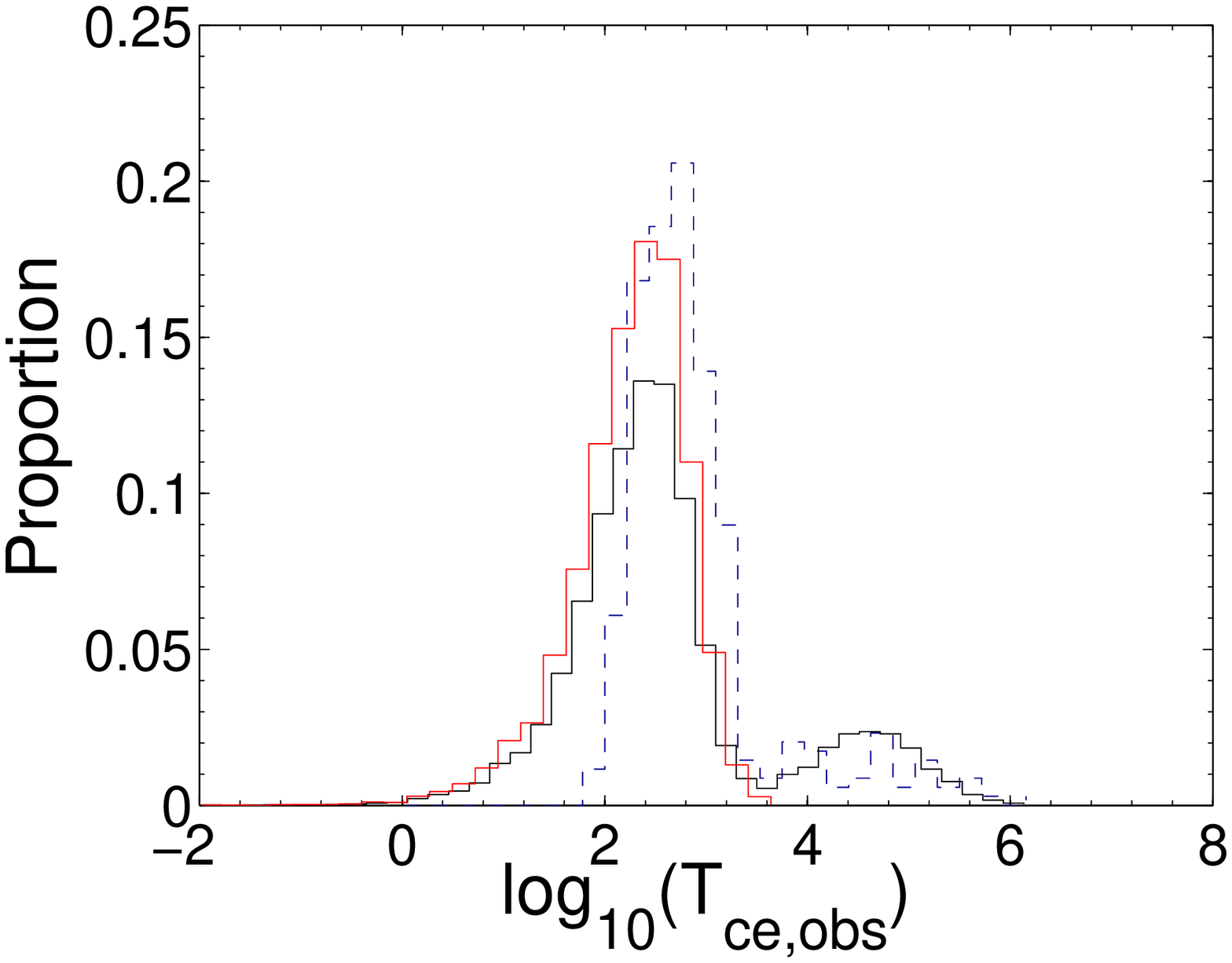}}\\
    \subfigure[]{
    \includegraphics[width=2.0in]{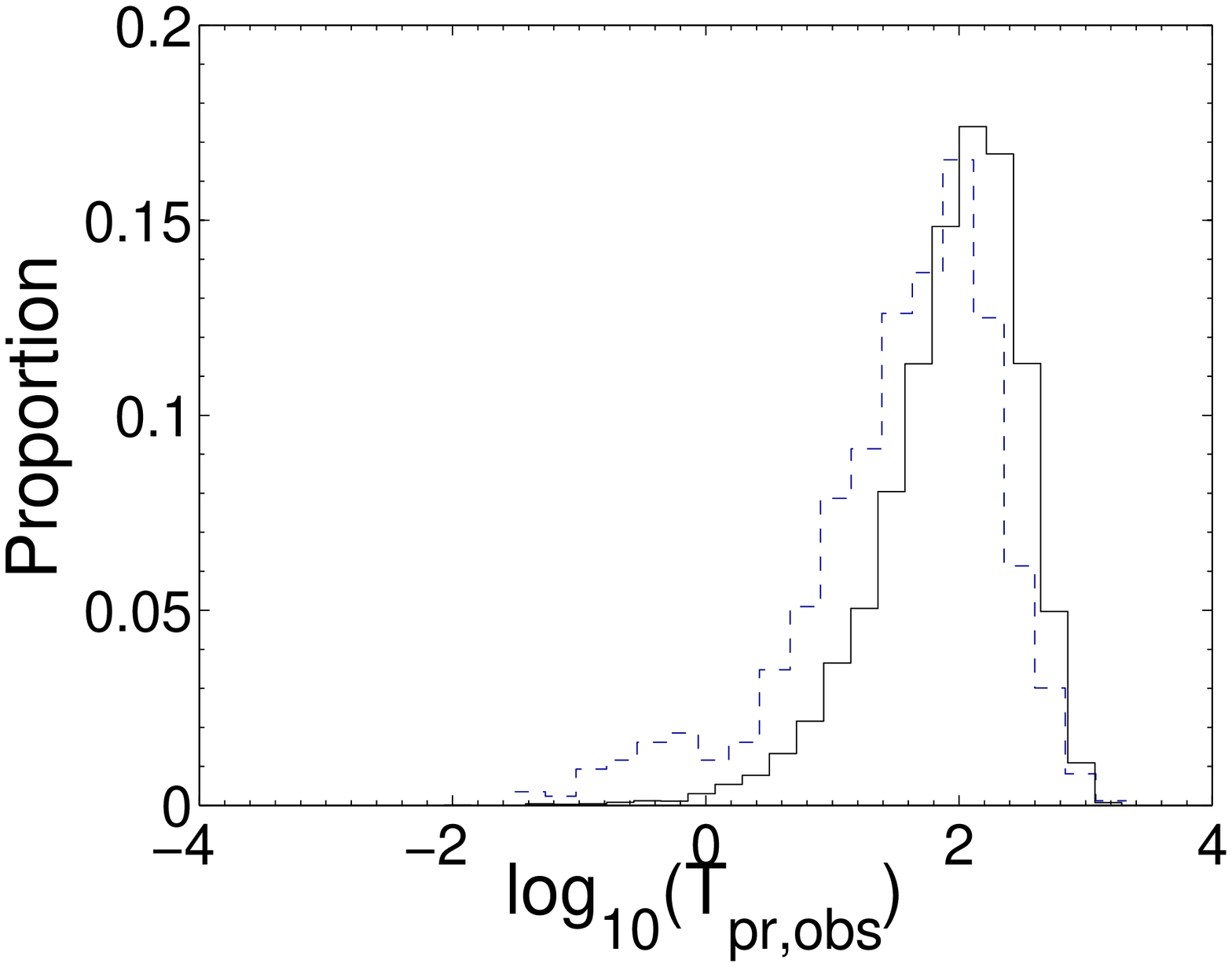}}
    \subfigure[]{
    \includegraphics[width=2.0in]{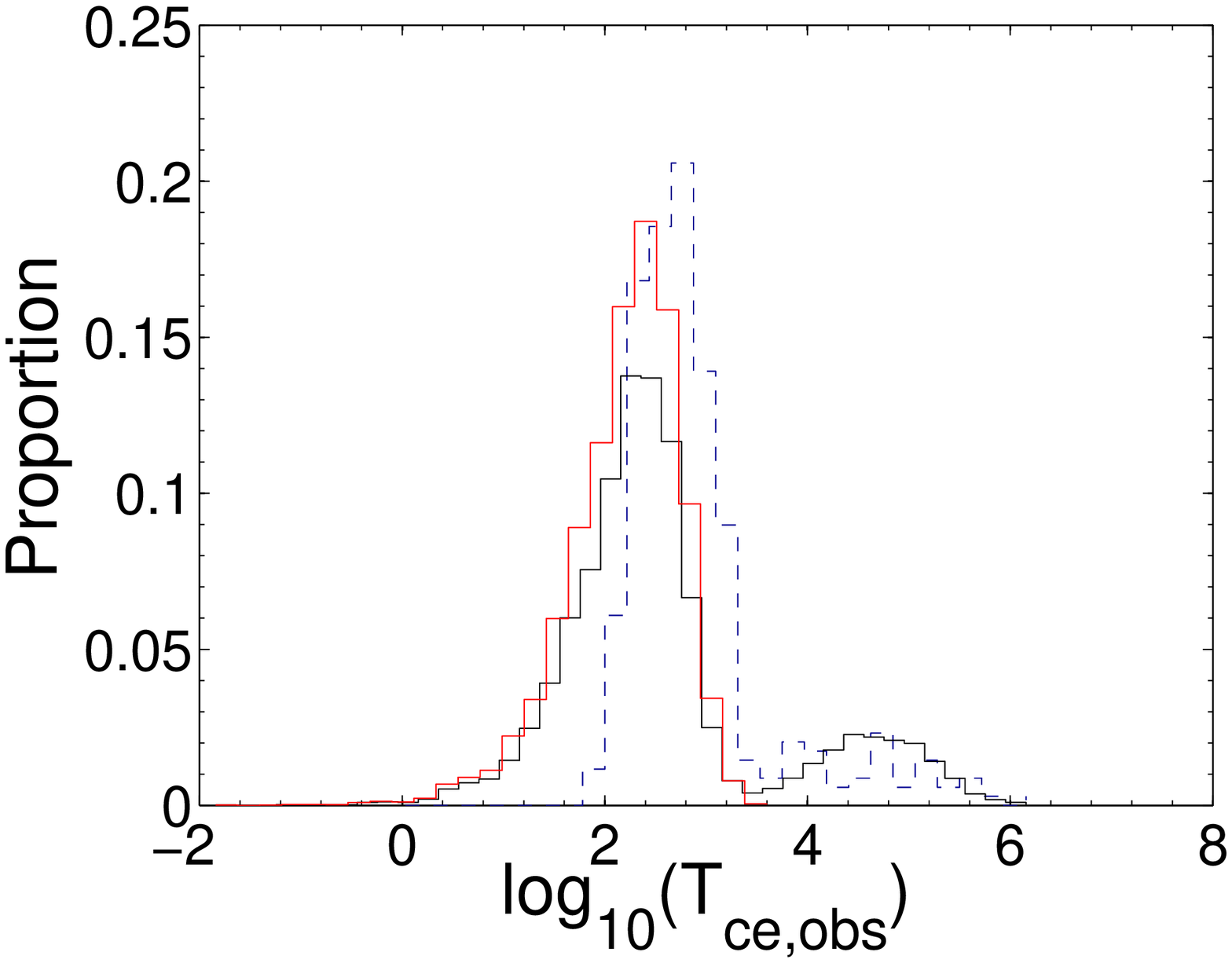}}
      \caption{An example of a $T_{\rm ce}$ ($T_{\rm ej}$) distribution that gives reasonable results for both $T_{90}$ ($T_{\rm pr,obs}$) and $t_{\rm burst}$ ($T_{\rm ce,obs}$).  (a) distribution of $T_{\rm ej}$ in log space; (b) distribution of first component for $T_{\rm ej}$ in linear space; $T_{\rm pr,obs}$ and $T_{\rm ce,obs}$ distributions for (c-d) RAN-type $\g$ initial condition, (e-f) PL-type $\g$ initial condition, and (g-h) GAUSS-type $\g$ initial condition. Solid and dash lines represent simulation results and observational data respectively. Red solid lines are for the results without considering the second component for $T_{\rm ej}$.}
            \end{figure}

\end{document}